\documentclass[journal]{IEEEtran}

\usepackage{lineno,hyperref}
\usepackage{bm}
\usepackage{mathrsfs}
\usepackage{subfigure} 
\usepackage{amssymb}
\usepackage{amsmath}
\usepackage{graphicx}
\usepackage{epstopdf}
\usepackage{multirow}
\usepackage{textcomp,booktabs}
\usepackage[usenames,dvipsnames]{color}
\usepackage[ruled,linesnumbered]{algorithm2e}
\usepackage{colortbl}
\definecolor{mygray}{gray}{.9}
\usepackage{cite}
\usepackage[numbers,sort&compress]{natbib}
\modulolinenumbers[5]

\ifCLASSINFOpdf

\else

\fi

\begin{document}
	\newcommand{\tabincell}[2]{\begin{tabular}{@{}#1@{}}#2\end{tabular}}
	
	\title{Gaussian-Trigonometric Functional Link Artificial Neural Network: Design and Analysis}
	
	\author{Jie Wang, Lu Lu,~\IEEEmembership{Senior Member,~IEEE,} Yi Yu,~\IEEEmembership{Senior Member,~IEEE,} Xiaodong Li, Chengshi Zheng,~\IEEEmembership{Senior Member,~IEEE,} and Rodrigo C. de Lamare,~\IEEEmembership{Fellow,~IEEE}
		
		\thanks{Manuscript received March 26, 2026. The work was supported by the NSFC under Grants 62371319 and 62471412, and the NSFSC under Grant 2026NSFSC0451.}
		\thanks{Jie Wang and Lu Lu are with the College of Electronics and Information Engineering, Sichuan University, Chengdu 610065, China (e-mail: wang\_jie\_1999@163.com and lulu19900303@126.com).}
		\thanks{Yi Yu is with the School of Information and Control Engineering, Southwest University of Science and Technology, Mianyang 621010, China, and also with the Hydrogen Energy and Multi-Energy Complementary Microgrid Engineering Technology Research Center of Sichuan Province, Mianyang 621010, China (e-mail: yuyi\_xyuan@163.com).}
		\thanks{Xiaodong Li and Chengshi Zheng are with the Key Laboratory of Noise and Vibration Research, Institute of Acoustics, Chinese Academy of Sciences, Beijing 100190, China, and also with the University of Chinese Academy of Sciences, Beijing 100049, China (e-mail: lxd@mail.ioa.ac.cn and cszheng@mail.ioa.ac.cn).}
		\thanks{Rodrigo C. de Lamare is with the Centre for Telecommunications Studies, Department of Electrical Engineering, Pontifical Catholic University of Rio de Janeiro (PUC-Rio), Rio de Janeiro 22451-900, Brazil (e-mail: delamare@puc-rio.br).}
		\thanks{Corresponding authors: Lu Lu and Chengshi Zheng.}}

	\markboth{}
	{Shell \MakeLowercase{\textit{et al.}}: Bare Demo of IEEEtran.cls for IEEE Journals}

	\maketitle
	\begin{abstract}
		This paper proposes a Gaussian function-based trigonometric functional link artificial neural network (GTFLN) filter for linear-in-the-parameters nonlinear filtering. Compared with the adaptive exponential TFLN (AETFLN) filter, the GTFLN filter provides smooth and localized basis functions with reduced computational complexity, where modeling advantages are theoretically established through the smoothness, reproducing kernel Hilbert space, approximation error, and operator theory properties. To maximize the modeling performance, an optimized scaling parameter for the GTFLN filter is derived, yielding the optimized GTFLN (OGTFLN) filter. Least mean square (LMS) adaptation is applied to the GTFLN and OGTFLN filters for nonlinear system identification, resulting in the GTFLMS and OGTFLMS algorithms, respectively. Moreover, the theoretical steady-state excess mean-square error of the GTFLN filter is analyzed. Simulations validate the effectiveness of the theoretical analysis and demonstrate the improved performance of the GTFLN and OGTFLN filters over the linear-in-the-parameters benchmarks in nonlinear system identification and nonlinear acoustic echo cancellation. Based on the GTFLN filter, the filtered-g LMS (FgLMS) algorithm is proposed for nonlinear active noise control. Simulations demonstrate improved stability and noise reduction performance compared to the benchmarks. 
	\end{abstract}
	\section{Introduction}
	\label{sec1}
	\IEEEPARstart{S}{y}stem identification is a fundamental problem in adaptive filtering. In conventional linear system identification (LSI), the least mean square (LMS)-based algorithms have been extensively investigated \cite{sayed2003fundamentals}. However, their performance is limited by nonlinearities in practical systems \cite{nguyen2012adaptive,liu2011kernel}. To address the limitations, nonlinear filters have been widely investigated for nonlinear system identification (NSI), nonlinear acoustic echo cancellation (NAEC), and nonlinear active noise control (NANC) \cite{10951113,YIN2024110887,11363386,sun2023adaptive,9788014}. 
	\subsection{Prior and Related Work}
	The linear-in-the-parameters (LIP) filters have been effectively adopted to characterize nonlinearities \cite{9893846}. Supported by the \textit{Stone-Weierstrass theorem}, Volterra-based models can approximate nonlinear dynamics via higher-order and cross-term expansions \cite{8786251}, where the second-order Volterra (SOV) filter achieves excellent performance in NSI \cite{8786251}, but the number of coefficients grows exponentially with filter order, incurring the curse of dimensionality. To alleviate this problem, functional link artificial neural networks (FLNs) have been explored, such as those based on Legendre \cite{carini2015legendre} and Chebyshev \cite{CARINI201624} polynomials, which achieve robust modeling capability with reduced computational complexity \cite{sicuranza2011generalized}.
	
	As a  representative LIP filter, the trigonometric FLN (TFLN) filter mapped inputs to a high-dimensional feature space through trigonometric expansions. However, its performance degraded in complex nonlinear systems \cite{patel2016design,6488741}. To improve modeling capability, the generalized TFLN (GeTFLN) filter introduced cross-terms between input samples and their delayed expansions, achieving modeling accuracy comparable to that of the high-order Volterra filter but at greater computational cost than the conventional TFLN \cite{sicuranza2011generalized}. In \cite{patel2016design}, the adaptive exponential TFLN (AETFLN) filter has augmented the trigonometric expansion with an adaptive exponential function (AEF) to improve estimation accuracy. 
	
	One typical application of LIP filters lies in NANC. In NANC, TFLN-based filters mitigate the performance degradation of conventional linear filters caused by nonlinear distortions in ANC systems \cite{tan2001adaptive,das2004active,patel2016design,9444125,sicuranza2011generalized}. The aforementioned SOV, TFLN, GeTFLN, and AETFLN filters can be combined with LMS-type adaptation and extended to NANC, yielding the SOV filtered-x LMS (VFxLMS) \cite{tan2001adaptive}, filtered-s LMS (FsLMS) \cite{das2004active}, generalized FsLMS (GFsLMS) \cite{sicuranza2011generalized}, and adaptive exponential FsLMS (AEFsLMS) \cite{patel2016design} algorithms, respectively. The properties of these NANC algorithms mirror those of their counterparts in NSI. These NANC algorithms inherit the main characteristics of their NSI counterparts.
	
	Alternatively, kernel adaptive filters (KAFs) leverage Mercer’s theorem to map inputs into a reproducing kernel Hilbert space (RKHS), providing strong nonlinear modeling capability \cite{liu2011kernel}, but their computational complexity typically grows with the dictionary size \cite{8786251}. Spline adaptive filters offer fixed-dimensional nonlinear expansions via piecewise polynomials, achieving improved performance with reduced complexity \cite{7305809}. However, both classes of methods, when viewed within the fixed-dimensional LIP framework, exhibit certain limitations: KAFs  rely on implicit mappings without an explicit fixed-dimensional feature expansion, restricting their deployment flexibility in resource-constrained scenarios, while conventional TFLN filters retain fixed dimensionality but their trigonometric basis functions (TBFs)-only expansions poorly capture localized variations in complex nonlinear systems \cite{patel2016design}. The AETFLN filter introduces an adaptive exponential envelope into the TBFs to improve localization \cite{patel2016design}, yet its envelope is nondifferentiable at the origin, which hinders stability conditions and high-order approximation accuracy.

	\subsection{Contributions}
	Motivated by these considerations, this work proposes the Gaussian function (GF)-based TFLN (GTFLN) filter, a fixed-dimensional nonlinear filter that combines a smooth localized envelope with kernel-level approximation capability. In contrast to the implicit dictionary-growing mechanism of KAFs, the GTFLN filter employs an explicit fixed-dimensional GF as a localization envelope for the TBFs. Moreover, from a Fourier-domain perspective, the GF exhibits exponential spectral decay and thus belongs to a much smoother class of RKHSs than the exponential envelope used in the AETFLN filter, whose Fourier transform decays only algebraically. This property endows the GTFLN filter with RKHS-level modeling capability comparable to that of kernel methods, while preserving a fixed-dimensional LIP structure and low computational complexity, thereby achieving a favorable tradeoff between modeling accuracy and deployment efficiency. An optimized scaling parameter is derived to further improve the nonlinear modeling performance, leading to the optimized GTFLN (OGTFLN) filter. Based on the GTFLN and OGTFLN architectures, two LMS-type algorithms, referred to as the GTFLMS and OGTFLMS algorithms, respectively, are developed with improved modeling accuracy for NSI. A theoretical expression for the steady-state excess mean-square error (EMSE) is derived. Moreover, the GTFLN-based filtered-g LMS (FgLMS) algorithm is developed for NANC, and the GTFLN framework is extended to NAEC. The main contributions of this work are summarized as follows.
	
	1) \textit{Novel Nonlinear Filters:} The GTFLN filter is proposed to overcome the limited localization capability of the TFLN filters. Unlike the AETFLN filter, the GTFLN filter is infinitely differentiable at the origin and incorporates the scaling parameter that controls the degree of localization, thereby improving its ability to model complex nonlinear systems. Using the energy conservation argument (ECA), an asymptotically optimal update for the scaling parameter is derived, yielding the OGTFLN filter. Building on the GTFLN framework, the GTFLMS and OGTFLMS algorithms for NSI are developed and the GTFLN filter is applied to NAEC. Moreover, the GTFLN-based FgLMS algorithm is developed for NANC.
	
	2) \textit{Performance Analysis:} The effects of the expansion order, tap length, and scaling parameter on the best approximation error of the GTFLN filter are analyzed and  the theoretical expression for the steady-state EMSE of the GTFLMS algorithm is derived by using the ECA and Taylor expansion. In addition, the superior modeling capability of the GTFLN filter relative to the  AETFLN filter is analyzed from the perspectives of smoothness, RKHS, approximation error, and operator theory. 
	
	3) \textit{Performance Validation:} Extensive simulations validate the theoretical steady-state EMSE analysis of the GTFLMS algorithm. The GTFLN and OGTFLN filters achieve higher estimation accuracy and lower steady-state errors than the LIP benchmarks in NSI and NAEC applications. Moreover, the FgLMS algorithm provides improved noise reduction performance and stability under various NANC scenarios. 
	
	The remainder of this work is organized as follows. Section \ref{PK} presents the GTFLN and OGTFLN filters. Section \ref{performance_analysis} provides the theoretical analysis. In Section \ref{simulations}, simulations validate the theory, demonstrate the advantages of GTFLN and OGTFLN filters for NSI and the applicability of GTFLN filter to NAEC, and evaluate the noise reduction of FgLMS for NANC. Finally, Section \ref{conclusion} concludes this paper.

	\section{Proposed GTFLN and OGTFLN Filters}
		\label{PK}
		In this section, we detail the proposed GTFLN and OGTFLN filters.
		\subsection{GTFLN Filter}
		The Gaussian function provides a smooth and localized representation whose degree of localization can be controlled by a scaling parameter. In the proposed GTFLN filter, the Gaussian function is not used as a reproducing kernel to induce an implicit mapping into an RKHS. Instead, it serves as an explicit zero-centered window that multiplies the TBFs. This construction preserves the fixed-dimensional LIP structure of the conventional TFLN filter while improving its ability to capture localized variations in nonlinear systems. To control the degree of localization, the scaling parameter $\gamma>0$ is introduced. The GF is defined as
		\begin{equation}
			\mathcal{G}(n)=\exp\left[-\gamma x^2(n)\right]\text{.}
			\label{077}
		\end{equation}
		\emph{Remark 1:} The GF in (\ref{077}) is a zero-centered member of the Gaussian radial basis function (RBF) family. Finite linear combinations of appropriately translated Gaussian RBFs possess the universal approximation property on compact domains and uniformly approximate any continuous nonlinear mapping with arbitrary accuracy \cite{392253}. Moreover, the corresponding translation-invariant Gaussian kernel is positive definite and uniquely induces an RKHS, providing a rigorous analytic framework for approximation and convergence analysis \cite{4609923}.        
		
		The proposed GTFLN filter incorporates the GF into the TFLN filter to capture the local characteristics of the input signal and enhance the modeling of nonlinear systems.
		\begin{figure}[!htbp]
			\centering
			\includegraphics[scale=0.02]{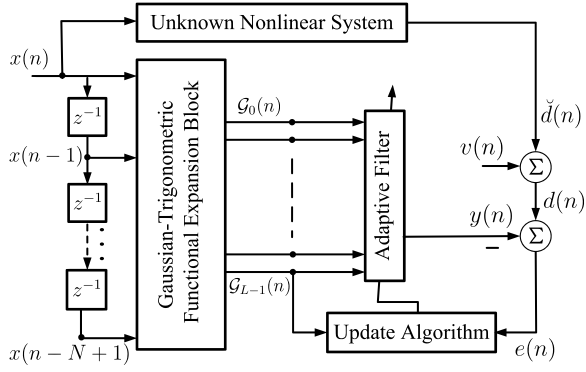}
			\caption{Block diagram of the GTFLN adopted by the nonlinear filter.}
			\label{Fig1}
		\end{figure}
		Fig. \ref{Fig1} plots the GTFLN-based nonlinear adaptive filter for the NSI task. The filter consists of a nonlinear functional expansion followed by the adaptive filter.   
		Specifically, the input signal is first mapped into a nonlinear feature space, and then the features are weighted by adaptive coefficients and summed to produce the filter output \cite{patel2020convergence}.
	The $N$-tap input vector is defined as $\bm x(n) = \left[x(n)\text{,}\; x(n-1)\text{,}\;\cdots\text{,}\;x(n-N+1)\right]^\mathrm{T}\in\mathbb{R}^{N\times 1}$, where $\left(\cdot\right)^\mathrm{T}$ denotes the transpose operator. The expanded input vector generated by the GTFLN filter is given by
	\begin{equation}
		\mathcal{\bm A}(n) = \left[1 \text{,}\; \bm x_{s_0}^\mathrm{T}(n)\text{,}\;\bm x_{s_1}^\mathrm{T}(n)\text{,}\;\cdots \text{,}\; \bm x_{s_{N-1}}^\mathrm{T}(n)\right]^\mathrm{T}\in\mathbb{R}^{L\times 1}\text{,}
		\label{004}
	\end{equation}
	where $\bm x_{s_j}(n) = \bigl[ x(n-j)\text{,}\allowbreak\; \mathcal{G}_{1\text{,}\,j}(n)\text{,}\allowbreak\; \mathcal{Q}_{1\text{,}\,j}(n)\text{,}\allowbreak\; \cdots\text{,}\allowbreak\;     \mathcal{G}_{i\text{,}\,j}(n)\text{,}\allowbreak\; \mathcal{Q}_{i\text{,}\,j}(n)\text{,}\allowbreak\; \cdots\text{,}\allowbreak\; \mathcal{G}_{\mathcal{B}\text{,}\,j}(n)\text{,}\allowbreak\; \mathcal{Q}_{\mathcal{B}\text{,}\,j}(n) \bigr]^{\mathrm T}$ with $i = 1\text{,}\,2\text{,}\,\cdots\text{,}\,\mathcal{B}$ and $j = 0\text{,}\,1\text{,}\,\cdots\text{,}\,N-1$, $\mathcal{B}$ denotes the expansion order, and $L = N(2\mathcal{B} + 1) + 1$. $\mathcal{G}_{i\text{,}j}(n)$ and $\mathcal{Q}_{i\text{,}j}(n)$ denote the GF-based TBFs for the GTFLN filter and are expressed as
	\begin{equation}
		\mathcal{G}_{i\text{,}\,j}(n) = \exp\left[-\gamma x^2(n-j)\right]\sin\left[i\pi x(n-j)\right]
		\label{r01}
	\end{equation}
	and
	\begin{equation}
		\mathcal{Q}_{i\text{,}\,j}(n) = \exp\left[-\gamma x^2(n-j)\right]\cos\left[i\pi x(n-j)\right]
		\label{r02}\text{.}
	\end{equation}
	\emph{Remark 2:} The GTFLN filter is different from the dictionary-based KAF. In the GTFLN filter, the GF serves as the zero-centered envelope for the TBFs and provides a localized representation of the input, while the trigonometric components capture periodic features at different harmonic orders, as defined in (\ref{r01}) and (\ref{r02}).  Moreover, the dimension of the GTFLN filter is determined by the tap length and expansion order and remains fixed as the number of input samples increases.
	
	The output of the nonlinear adaptive filter is obtained by multiplying  $\mathcal{\bm A}(n)$ with the adaptive weight vector $\bm \omega(n)= \left[\omega_0(n)\text{,}\;\omega_1(n)\text{,}\;\cdots\text{,}\;\omega_{L-1}(n)\right]^\mathrm{T}\in\mathbb{R}^{L\times 1}$ and is expressed as
	\begin{equation}
		y(n) =\bm \omega^\mathrm{T}(n)\mathcal{\bm A}(n)\text{.}
		\label{005}
	\end{equation}
Then, the error signal of the GTFLN filter is given by 
\begin{equation}
	e(n) = \breve{d}(n) + v(n) - y(n) = d(n) - y(n)\text{,}
	\label{007}
\end{equation}
where $d(n)=\breve{d}(n) + v(n)$ denotes the desired signal, $v(n)$ is a zero-mean white Gaussian noise (WGN) with a variance $\sigma_v^2$, and $\breve{d}(n) = \bm \omega_o^\mathrm{T}\mathcal{\bm A}(n)$ denotes the system output.

The weight coefficients for the GTFLN filter are adapted by minimizing the mean-square error (MSE) cost function 
\begin{equation}
	\mathcal{J}(n) = \mathit{E}\left\{e^2(n)\right\}\text{,}
	\label{006}
\end{equation}
where $\mathit{E}\left\{\cdot\right\}$ denotes the expectation.	Under the ergodicity assumption, the MSE can be estimated by time-averaging the instantaneous squared error and as such, we have $\mathcal{J}(n) = e^2(n) \approx \mathit{E}\left\{e^2(n)\right\}$ \cite{sayed2003fundamentals}. Using the stochastic gradient descent approach, the weight for the GTFLN filter is updated by
\begin{equation}
	\bm{\omega}(n+1) = \bm\omega(n) - \frac{\mu}{2}\nabla_{\bm\omega}\mathcal{J}(n)\text{,}
	\label{008}
\end{equation}
where $\mu$ is the step size and $\nabla_{\bm\omega}\mathcal{J}(n)$ is the instantaneous estimate of the MSE gradient. Using (\ref{005}) and (\ref{007}), the update rule of the GTFLN filter can be rewritten as
\begin{equation}
	\bm \omega(n+1) = \bm \omega(n) + \mu e(n)\mathcal{\bm A}(n)\text{.}
	\label{009}
\end{equation}
One can see that (\ref{009}) corresponds to the GTFLMS algorithm.

\subsection{Improved Modeling Capability of GTFLN over AETFLN}
	As an extension of the conventional TFLN filters, the AETFLN filter incorporates an exponentially varying envelope into the TBFs to enhance nonlinear modeling performance. Similar to the GTFLN filter, the AETFLN filter performs a nonlinear expansion of the $N$-tap input vector $\bm x(n)$. The expanded input vector $\mathcal{\bm X}_{A}(n)\in\mathbb{R}^{L\times 1}$ is given by \cite{patel2016design}
	\begin{equation}
		\mathcal{\bm X}_{A}(n) = \left[1 \text{,}\; \bm x_{a\text{,}\,0}^\mathrm{T}(n)\text{,}\;\bm x_{a\text{,}\,1}^\mathrm{T}(n)\text{,}\;\cdots \text{,}\; \bm x_{a\text{,}\,{N-1}}^\mathrm{T}(n)\right]^\mathrm{T}\text{,}
		\label{101}
	\end{equation}
	where $\bm x_{a\text{,}\,j}(n) = \bigl[ x(n-j)\text{,}\allowbreak\; \mathcal{U}_{s\text{,}\,{1j}}(n)\text{,}\allowbreak\; \mathcal{U}_{c\text{,}\,{1j}}(n)\text{,}\allowbreak\; \cdots\text{,}\allowbreak\; \mathcal{U}_{s\text{,}\,{ij}}(n)\text{,}\allowbreak\; 
	\mathcal{U}_{c\text{,}\,{ij}}(n)\text{,}\allowbreak\; \cdots\text{,}\allowbreak\; \mathcal{U}_{s\text{,}\,{\mathcal{B}j}}(n)\text{,}\allowbreak\; 
	\mathcal{U}_{c\text{,}\,{\mathcal{B}j}}(n) \bigr]^{\mathrm T}$. $\mathcal{U}_{s\text{,}\,{ij}}(n)$ and $\mathcal{U}_{c\text{,}\,{ij}}(n)$ denote the exponentially varying TBFs for the AETFLN filter and are expressed as
	\begin{equation}
		\mathcal{U}_{s\text{,}\,{ij}}(n)=e^{-a(n)\left|x(n-j)\right|}\sin\left[i\pi x(n-j)\right] \label{701}
	\end{equation}
	and
	\begin{equation}
		\mathcal{U}_{c\text{,}\,{ij}}(n)=e^{-a(n)\left|x(n-j)\right|}\cos\left[i\pi x(n-j)\right]\label{702}\text{,}
	\end{equation} 
	where $a(n)$ is the adaptive exponential parameter and $\left|\cdot\right|$ stands for the absolute value operation.  By following a similar procedure with (\ref{006}) and (\ref{008}), the LMS-type adaptation of the AETFLN filter with respect to the weight vector and adaptive exponential parameter can be obtained. As shown in (\ref{701}) and (\ref{702}), the AETFLN filter uses the AEF $\varsigma\left[x(n)\right]=\exp\left[-a(n)|x(n)|\right]$ to generate localized trigonometric features for nonlinear modeling. 
	
	\emph{Remark 3:} Appendices \ref{appen1}-\ref{theory4} provide a theoretical comparison between the GTFLN and AETFLN filters. The GF $\mathcal{G}[x(n)]=\exp[-\gamma x^2(n)]$ used in the GTFLN filter is compared with the AEF $\varsigma[x(n)]=\exp[-a(n)\vert x(n)\vert]$ used in the AETFLN filter in terms of smoothness, membership in Mat\'{e}rn RKHSs, approximation error, and operator theory properties. These analyses establish the theoretical advantages of the GTFLN filter over the AETFLN filter, and the performance improvement in nonlinear modeling is validated through simulations.

1) \textit{Smoothness:} The GF satisfies $\mathcal{G}\left[x(n)\right]\in\mathit{C}^\infty(\mathbb{R})$ and is infinitely differentiable with respect to its input. In contrast, the AEF satisfies $\varsigma\left[x(n)\right]\in\mathit{C}^{0}(\mathbb{R})$ but $\varsigma\left[x(n)\right]\notin\mathit{C}^{1}(\mathbb{R})$ because its first derivative is discontinuous at the origin.

2) \textit{RKHS:} The Fourier transform of $\mathcal{G}[x(n)]$ exhibits exponential decay, implying membership in the RKHS for every $\upsilon>0$. In contrast, the Fourier transform of $\varsigma[x(n)]$ decays algebraically, restricting its RKHS membership to $\upsilon<1$.

3) \textit{Approximation Error:} The approximation error of $\mathcal{G}[x(n)]$ converges super-exponentially at a rate of $\mathcal{O}(1/\sqrt{k!})$, whereas that of $\varsigma[x(n)]$ converges algebraically at a rate of $\mathcal{O}(1/k)$. Thus, the GF provides a faster error decay.

4) \textit{Operator Theory:} The GF $\mathcal{G}[x(n)]$ is proportional to the ground-state eigenfunction of the harmonic oscillator. In contrast, the AEF $\varsigma[x(n)]$ corresponds to a bound state of a singular delta potential, whose distributional second derivative contains a Dirac delta term at the origin, consistent with the discontinuity of its first derivative.

\subsection{OGTFLN Filter}
The fixed Gaussian function $\exp\left[-x^2(n)\right]$ may lack sufficient flexibility for accurate nonlinear modeling. To address this limitation, the scaling parameter $\gamma$ is introduced to control the degree of localization. This subsection derives the optimized scaling parameter $\gamma_o$ to improve the nonlinear modeling accuracy and overall performance of the GTFLN filter.

Taking $\mathit{L}_2$-norm and expectations on both sides of (\ref{008}) yields
\begin{equation}
	\begin{aligned}
		\mathit{E}\left\{\left\|\bm\omega(n+1)\right\|^2\right\}\!
		&=\mathit{E}\left\{\left\|\bm\omega(n)\right\|^2\right\} + \mu^2\mathit{E}\left\{e^2(n)\left\|\mathcal{\bm A}(n)\right\|^2\right\}\\
		&\quad+ 2\mu \mathit{E}\left\{e(n)\bm \omega^\mathrm{T}(n)\mathcal{\bm A}(n)\right\}\text{.}
		\label{010}
	\end{aligned}
\end{equation}
Assume that $\lim\limits_{n\to\infty}\mathit{E}\left\{\left\|\bm\omega(n+1)\right\|^2\right\} = \lim\limits_{n\to\infty}\mathit{E}\left\{\left\|\bm\omega(n)\right\|^2\right\}$ at steady state \cite{sayed2003fundamentals}. Such an assumption is widely utilized as the convergence condition and as such, we have
\begin{equation}
	2\mu \mathit{E}\left\{e(n)\bm \omega^\mathrm{T}(n)\mathcal{\bm A}(n)\right\}\\
	+ \mu^2\mathit{E}\left\{e^2(n)\left\|\mathcal{\bm A}(n)\right\|^2\right\}=0\text{.}
	\label{011}
\end{equation}
Introducing $\gamma_o$ into the GTFLN filter, substituting (\ref{005}) into (\ref{007}), and applying the convergence condition yield
\begin{equation}
	e(n)=\bm \omega_o^\mathrm{T}\mathcal{\bm A}_o(n) + v(n) - \bm \omega^\mathrm{T}(n)\mathcal{\bm A}(n) \approx \bm \omega_o^\mathrm{T}\bm \Gamma(n) + v(n)
	\text{,}
	\label{012}
\end{equation}
where $\bm\Gamma(n) = \mathcal{\bm A}_o(n) - \mathcal{\bm A}(n)$ and $\mathcal{\bm A}_o(n)$ has exactly the same ordering as $\mathcal{A}(n)$ in (\ref{004}), with $\gamma$ replaced by $\gamma_o$. 
To obtain $\gamma_o$, the Taylor series expansion of $f(x)$ is introduced as \cite{4609923}
\begin{equation}
	f(x) = f(x_o) + f'(x_o)(x-x_o) + \mathrm{R}\left[(x-x_o)\right]\text{,}
	\label{016}
\end{equation}
where $\mathrm{R}\left[\cdot\right]$ represents the Taylor remainder term. Two terms $\mathcal{S}_{(j\text{,}i)}(n)$ and $\mathcal{C}_{(j\text{,}i)}(n)$ are defined as
\begin{equation}
	\mathcal{S}_{(j\text{,}i)}(n) = \left[e^{-\gamma_o x^2(n-j)} - e^{-\gamma x^2(n-j)}\right]\sin\left[i\pi x(n-j)\right]
	\label{015}
\end{equation}
and
\begin{equation}
	\mathcal{C}_{(j\text{,}i)}(n) = \left[e^{-\gamma_o x^2(n-j)} - e^{-\gamma x^2(n-j)}\right]\cos\left[i\pi x(n-j)\right]\text{.}
	\label{014}
\end{equation}

By applying a first-order Taylor series expansion to $\mathcal{S}_{(j\text{,}i)}(n)$ and neglecting the quadratic and above terms concerning $-(\gamma_o - \gamma)x^2(n-j)$, $\mathcal{S}_{(j\text{,}i)}(n)$ can be calculated as
\begin{equation}
	\mathcal{S}_{(j\text{,}i)}(n)\approx-\varphi x^2(n-j)e^{-\gamma x^2(n-j)}\sin\left[i\pi x(n-j)\right]\text{,}
	\label{017}
\end{equation}
where $\varphi = \gamma_o - \gamma$. Similarly, reusing the Taylor expansion for $\mathcal{C}_{(j\text{,}i)}(n)$ yields
\begin{equation}
	\mathcal{C}_{(j\text{,}i)}(n)
	\approx-\varphi x^2(n-j)e^{-\gamma x^2(n-j)}\cos\left[i\pi x(n-j)\right]\text{.}
	\label{019}
\end{equation}
Substituting (\ref{015})-(\ref{019}) into $\bm\Gamma(n)$ results in
\begin{equation}
	\bm \Gamma(n)
	=-\varphi\bm \Omega(n)\text{,}
	\label{020}
\end{equation}
where $\bm \Omega(n) =[0\text{,}\; 0\text{,}\; x^2(n)\mathcal{G}_{1\text{,}\,0}(n) \text{,}\;\cdots\text{,}\; x^2(n-N+1)\mathcal{Q}_{\mathcal{B}\text{,}\,N-1}(n)]^\mathrm{T}$. Given that the weight vector $\bm\omega_o$ of the unknown system is unavailable in practice, $\bm \omega(n)$ is assumed to asymptotically reach $\bm \omega_o$ under the slow-adaptation assumption \cite{sayed2003fundamentals}. Using (\ref{020}) and (\ref{012}), the error can be expressed as
\begin{equation}
	e(n) = -\varphi\bm \omega^\mathrm{T}(n)\bm \Omega(n) + v(n)\text{.}
	\label{022}
\end{equation} 
Substituting (\ref{022}) into (\ref{011}) and utilizing the fact that $v(n)$ is independent of the expanded input vector $\mathcal{\bm A}(n)$, we have
\begin{equation}
	a(n)\varphi^2 + b(n)\varphi + c(n) = 0\text{,}
	\label{023}
\end{equation}
where $a(n)$, $b(n)$, and $c(n)$ are computed by
\begin{equation}
	a(n)=\mu\mathit{E}\left\{\left|\bm\omega^\mathrm{T}(n)\bm{\Omega}(n)\right|^2\right\}\mathit{E}\left\{\left\|\mathcal{\bm A}(n)\right\|^2\right\}\text{,}
	\label{024}
\end{equation}
\begin{equation}
	b(n) = -2\mathit{E}\left\{\bm \omega^\mathrm{T}(n)\bm \Omega(n)\bm\omega^\mathrm{T}(n)\mathcal{\bm A}(n)\right\}\text{,}
	\label{025}
\end{equation}
and
\begin{equation}
	c(n) = \mu\sigma_v^2\mathit{E}\left\{\left\|\mathcal{\bm A}(n)\right\|^2\right\}\text{.}
	\label{026}
\end{equation}

Utilizing the quadratic formula to solve (\ref{023}) with $\varphi(n) = \gamma_o(n) - \gamma$, one solution is given by
\begin{equation}
	\varphi_1(n) = \frac{-b(n)-\sqrt{b^2(n) - 4a(n)c(n)}}{2a(n)}\text{.}
	\label{027}
\end{equation}

\textit{Remark 4:} The solution $\varphi_2(n)= \frac{-b(n)+\sqrt{b^2(n) - 4a(n)c(n)}}{2a(n)}$ in (\ref{023}) is inadmissible for the following reasons. Define $\epsilon_1(n) = \mathit{E}\{\vert\bm\omega^\mathrm{T}(n)\bm{\Omega}(n)\vert^2\}\mathit{E}\{\Vert\mathcal{\bm A}(n)\Vert^2\}$, $\epsilon_2(n) =2\mathit{E}\{\bm \omega^\mathrm{T}(n)\bm \Omega(n)\bm\omega^\mathrm{T}(n)\mathcal{\bm A}(n)\}$, and $\epsilon_3(n) = \sigma_v^2\mathit{E}\{\Vert\mathcal{\bm A}(n)\Vert^2\}$, respectively.  Under the slow adaptation and moment separation assumptions, $\epsilon_1(n)$, $\epsilon_2(n)$, and $\epsilon_3(n)$ remain bounded with respect to $\mu$. Based on (\ref{004}) and (\ref{025}), we have $\epsilon_2(n)>0$, which is equivalent to $b(n)<0$. Before dividing by the common factor $\mu$, substituting (\ref{022}) into (\ref{011}) leads to $\mu^2 \epsilon_1(n)\varphi^2(n) - \mu\epsilon_2(n)\varphi(n) + \mu^2\epsilon_3(n) = 0$, where the three coefficients are of orders $\mathcal{O}(\mu^2)$, $\mathcal{O}(\mu)$, and $\mathcal{O}(\mu^2)$, respectively. Dividing the equation by $\mu$ yields (\ref{023}), where $a(n) = \mu\epsilon_1(n)$, $b(n) = -\epsilon_2(n)$, and $c(n) = \mu\epsilon_3(n)$, respectively. Consequently, both normalizations yield the same ratio with $4a(n)c(n)/b^2(n) = 4\mu^2\epsilon_1(n)\epsilon_3(n)/\epsilon_2^2(n)=\mathcal{O}(\mu^2)$. Hence, we have $4\mu^2\epsilon_1(n)\epsilon_3(n)<\epsilon_2^2(n)$ for a small step size and the discriminant is positive with $4a(n)c(n)\ll b^2(n)$. Meanwhile, let $\vartheta(n) = 4\mu^2\epsilon_1(n)\epsilon_3(n)/\epsilon_2^2(n)$. Using $\sqrt{1-\vartheta(n)} = 1-\vartheta(n)/2+\mathcal{O}\left[\vartheta^2(n)\right]$, the two solutions of (\ref{023}) satisfy $\varphi_1(n) = \frac{\epsilon_2(n) - \sqrt{\epsilon_2^2(n) - 4\mu^2\epsilon_1(n)\epsilon_3(n)}}{2\mu\epsilon_1(n)} = \frac{\mu\epsilon_3(n)}{\epsilon_2(n)}+ \mathcal{O}(\mu^3) = \mathcal{O}(\mu)$ and $\varphi_2(n) = \frac{\epsilon_2(n) + \sqrt{\epsilon_2^2(n) - 4\mu^2\epsilon_1(n)\epsilon_3(n)}}{2\mu\epsilon_1(n)}= \frac{\epsilon_2(n)}{\mu\epsilon_1(n)} - \frac{\mu\epsilon_3(n)}{\epsilon_2(n)}+ \mathcal{O}(\mu^3)= \mathcal{O}(\mu^{-1})$. In addition, the first-order expansions in (\ref{017}) and (\ref{019}) require $\left|\varphi(n)x^2(n-j)\right|\ll 1$. For bounded-amplitude inputs, $\varphi_1(n) = \mathcal{O}(\mu)$ is consistent with the local expansion condition as $\mu\to 0$. In contrast, $\varphi_2(n) = \mathcal{O}(\mu^{-1})$ grows as $1/\mu$ and contradicts the approximations underlying (\ref{017}) and (\ref{019}), and the solution $\varphi_2(n)$ becomes asymptotically unbounded as $\mu\to 0$, which may cause the GTFLN filter to diverge. Hence, the asymptotically optimal scaling parameter $\gamma_o(n)$ can be expressed as
\begin{equation}
	\gamma_o(n) = \gamma + \frac{-b(n)-\sqrt{b^2(n) - 4a(n)c(n)}}{2a(n)}\text{.}
	\label{028}
\end{equation} 
By substituting (\ref{028}) into (\ref{004}), the enhanced GTFLN filter is obtained and termed the GTFLN filter with optimized scaling parameter (OGTFLN). Similar to the GTFLMS algorithm, equations (\ref{004}), (\ref{009}), and (\ref{028}) form the OGTFLMS algorithm.

\emph{Remark 5:} When incorporated into the TFLN structure, the GF acts as a parameterized and input-dependent Gaussian envelope. Its width is controlled by $\gamma$ in the GTFLN filter, whereas it is adjusted iteratively toward $\gamma_o$ in the OGTFLN filter.

\subsection{Computational Complexity}
The computational complexities and memory requirements of the proposed GTFLN and OGTFLN filters in comparison to the existing nonlinear expansion structures, including the SOV \cite{1453790}, TFLN \cite{das2004active}, GeTFLN \cite{sicuranza2011generalized}, and AETFLN \cite{patel2016design} filters utilized as the benchmarks, are provided in Table \ref{Table01} in terms of the multiplication/division, addition/subtraction, $\sin\left(\cdot\right)/\cos\left(\cdot\right)$, $\exp\left(\cdot\right)$, square root operation, and absolute value operation. For an $N$-tap input vector, the GTFLN filter contains the following two parts.

1) \textit{Expansion of the Input Signal:} It requires $2\mathcal{B}N$ for $\sin\left(\cdot\right)/\cos\left(\cdot\right)$ calculations, $3\mathcal{B}N + 2N+1$ multiplications, and $N$ for $\exp\left(\cdot\right)$ calculations.

2) \textit{Updating the Filter Weights:} It requires $4\mathcal{B}N + 2N+3$ multiplications and $4\mathcal{B}N + 2N+2$ additions.

The OGTFLN filter includes an additional step 3) for computing $\gamma_o$ using the results obtained in 1) and 2).

3) \textit{Calculation of the Optimized Scaling Parameter $\gamma_o$:} It requires $8\mathcal{B}N + 4N+17$ multiplications, $6\mathcal{B}N + 3N + 3$ additions, and $1$ square root calculation.

The GeTFLN filter incorporates cross-terms, with $P$ denoting the number of diagonal terms in the cross-term expansion. From Table \ref{Table01}, the GTFLN filter has lower computational complexity than the AETFLN filter. However, the OGTFLN filter requires higher computational complexity due to the calculation of the optimized scaling parameter $\gamma_o$.
\begin{table*}[!htp]
	\small
	\centering
	\scriptsize
	\caption{Comparison of Computational Complexities and Memory Requirements For Nonlinear Filters Per Iteration}
	\label{Table01}
	\renewcommand{\arraystretch}{1.8}
	\doublerulesep=0.6pt
	\begin{tabular}{cccccccc}
		\hline\hline
		\textbf{filters} &\textbf{$\times/\div$} &\textbf{$+/-$} &{$\sin(\cdot)/\cos(\cdot)$} &{exp($\cdot$)} &{square root} &{absolute} &{Memory} \\ \hline
		\textbf{SOV}           &$\frac{3}{2}N^2+\frac{7}{2}N+1$  &$N^2 + 3N$ &$0$ &$0$ &$0$ &$0$ &$N^2+4N+4$ \\ \hline
		\textbf{TFLN}             &$5\mathcal{B}N+2N+1$ &$4\mathcal{B}N+2N$ &$2\mathcal{B}N$ &$0$ &$0$ &$0$ &$4\mathcal{B}N+2N+4$ \\ \hline
		\textbf{GeTFLN}				       &\begin{tabular}[c]{@{}l@{}}$3\mathcal{B}(P+1)(2N-P)$\\$-\mathcal{B}N+2N+1$\end{tabular}  &$2\mathcal{B}(2N-P)(P+1) + 2N$ &$2\mathcal{B}N$ &$0$ &$0$ &$0$ &\begin{tabular}[c]{@{}l@{}} $4\mathcal{B}(P+1)(2N-P)$\\$+2N+4$\end{tabular} \\ \hline
		\textbf{AETFLN}             &$11\mathcal{B}N+5N+7$ &$6\mathcal{B}N+3N+3$ &$2\mathcal{B}N$ &$N$ &$0$ &$N$ &$6\mathcal{B}N+5N+10$ \\ \hline
		\textbf{GTFLN}         &$7\mathcal{B}N+4N+4$ &$4\mathcal{B}N+2N+2$ &$2\mathcal{B}N$ &$N$ &$0$ &$0$ &$4\mathcal{B}N+3N+7$ \\ \hline
		\textbf{OGTFLN} &$15\mathcal{B}N+8N+21$ &$10\mathcal{B}N+5N+5$ &$2\mathcal{B}N$ &$N$ &$1$ &$0$ &$6\mathcal{B}N+4N+13$ \\ \hline\hline
	\end{tabular}
\end{table*}
\section{Performance Analysis}
\label{performance_analysis}
This section analyzes the performance of the GTFLN filter, including the effects of the expansion order $\mathcal{B}$, tap length $N$, and scaling parameter $\gamma$ on the finite-dimensional approximation error, and an analysis of the theoretical steady-state EMSE in the mean-square sense. 
	\subsection{Finite-Dimensional Approximation Error of GTFLN}
	Unlike dictionary-based KAFs, whose model order grows as new input samples are admitted into the dictionary, the GTFLN filter adopts a fixed-dimensional vector for fixed values of $\mathcal{B}$ and $N$. The parameter $\gamma$ in the GF controls the localization of trigonometric features. This subsection characterizes the effects of $\mathcal{B}$, $N$, and $\gamma$ on the best approximation error of the finite GTFLN dictionary. Let $\mathcal{T}_{\mathcal{B}\text{,}\,N\text{,}\,\gamma}$ denote the linear span of the finite GTFLN  vector $\mathcal{A}_{\mathcal{B}\text{,}\,N\text{,}\,\gamma}(n)$. For the system output $\breve{d}(n)$, the minimum mean-square approximation error under the input distribution is defined as 
	\begin{equation}
		\begin{aligned}
			\varepsilon^2_{\rm app} \left(\mathcal{B}\text{,}\,N\text{,}\,\gamma\right)
			&= \inf_{\bm \omega\in\mathbb{R}^L}\mathit{E}\left\{\left|\breve{d}(n) - \bm\omega^\mathrm{T}\mathcal{A}_{\mathcal{B}\text{,}\,N\text{,}\,\gamma}(n)\right|^2\right\}\\
			& = \mathit{E}\left\{\breve{d}^2(n)\right\} - \bm p_{\mathcal{B}\text{,}\,N\text{,}\,\gamma}^\mathrm{T}\bm R^\dagger_{\mathcal{B}\text{,}\,N\text{,}\,\gamma}\bm p_{\mathcal{B}\text{,}\,N\text{,}\,\gamma}\text{,}
			\label{901}
		\end{aligned}
	\end{equation}
	where $\bm R_{\mathcal{B}\text{,}\,N\text{,}\,\gamma} = \mathit{E}\left\{\mathcal{A}_{\mathcal{B}\text{,}\,N\text{,}\,\gamma}(n)\mathcal{A}_{\mathcal{B}\text{,}\,N\text{,}\,\gamma}^\mathrm{T}(n)\right\}$, $\bm p_{\mathcal{B}\text{,}\,N\text{,}\,\gamma} = \mathit{E}\left\{\mathcal{A}_{\mathcal{B}\text{,}\,N\text{,}\,\gamma}(n)\breve{d}(n)\right\}$, and $(\cdot)^\dagger$ denotes the Moore-Penrose inverse. From (\ref{901}), the approximation error depends not only on $\mathcal{B}$, $N$, and $\gamma$, but also on the input distribution.
	
	\emph{Theorem 1:} For $\gamma\geq0$, the best approximation error is nonincreasing with the expansion order $\mathcal{B}$ and tap length $N$
	\begin{equation}
		\varepsilon_{\rm app}\left(\mathcal{B}+1\text{,}\,N\text{,}\,\gamma\right)\leq	\varepsilon_{\rm app}\left(\mathcal{B}\text{,}\,N\text{,}\,\gamma\right) 
		\label{902}
	\end{equation}
	and
	\begin{equation}
		\varepsilon_{\rm app}\left(\mathcal{B}\text{,}\,N+1\text{,}\,\gamma\right)\leq	\varepsilon_{\rm app}\left(\mathcal{B}\text{,}\,N\text{,}\,\gamma\right)\text{.}
		\label{903}
	\end{equation}
	Suppose that the normalized input satisfies $\mathit{X}_j=x(n-j)\in[-1\text{,}\,1]$ and $\breve{d}(n)$ admits the decomposition $\breve{d}(n) = \sum_{j=0}^{N-1}f_j\left(\mathit{X}_j\right) + \alpha_N(n) + c$, where $f_j\left(\mathit{X}_j\right)$ denotes the centered univariate nonlinear contribution associated with the $j$th input tap and satisfies $\mathit{E}\{f_j\left(\mathit{X}_j\right)\}=0$, $c=\mathit{E}\{\breve{d}(n)\}$ denotes the constant offset, and $\alpha_N(n) = \breve{d}(n) - c - \sum_{j=0}^{N-1}f_j\left(\mathit{X}_j\right)$, which implies $\mathit{E}\left\{\alpha_N(n)\right\} = 0$. Note that for unbounded Gaussian inputs, additional tail-integrability assumptions are required. The residual $\alpha_N(n)$ contains the contribution of the omitted system memory and the nonadditive cross-tap interactions that cannot be represented by the additive dictionary. For each $f_j\left(\mathit{X}_j\right)$, we have $k_{j\text{,}\,\gamma}\left(\mathit{X}_j\right) = e^{\gamma \mathit{X}_j^2}[f_j\left(\mathit{X}_j\right)-  \theta_j -\delta_j \mathit{X}_j]$, where $\theta_j $ and $\delta_j$ denote the auxiliary constant and auxiliary linear coefficient associated with the $j$th univariate functional component and are defined as
	\begin{equation}
		\theta_j = \frac{\int_{-1}^{1}e^{\gamma t^2}\left[f_j\left(t\right)- \delta_j t\right]\mathrm{d}t}{\int_{-1}^{1}e^{\gamma t^2}\mathrm{d}t}\;\;\;\text{and}\;\;\;\delta_j=\frac{f_j(1) - f_j(-1)}{2}\text{.}
		\label{905}
	\end{equation}
	Eq. (\ref{905}) ensures that $k_{j\text{,}\,\gamma}(-1)=k_{j\text{,}\,\gamma}(1)$ and $\int_{-1}^{1}k_{j\text{,}\,\gamma}(t)\mathrm{d}t = 0$, respectively. Suppose that $k_{j\text{,}\,\gamma}$ has a periodic extension with period 2 in $H^s_{\rm per}([-1\text{,}\,1])$ for an integer $s\geq1$. For $s>1$, the corresponding endpoint compatibility conditions on the derivatives up to order $s-1$ are also assumed. Finally, assume that the marginal density $q_j$ of $x(n-j)$ exists and satisfies $0\leq q_j(t)\leq \mathcal{M}_j<\infty$ with $t\in[-1\text{,}\,1]$. Then, the approximation error is bounded by
	\begin{equation}
		\varepsilon_{\rm app}\left(\mathcal{B}\text{,}\,N\text{,}\,\gamma\right)
		\leq
		\left\|\alpha_N\right\|_{L^2(P)} + \frac{\sum_{j=0}^{N-1}\sqrt{\mathcal{M}_j}\left\|k_{j\text{,}\,\gamma}^{(s)}\right\|_{L^2(-1\text{,}\,1)}}{\left[\pi (\mathcal{B}+1)\right]^s}\text{.}
		\label{906}
	\end{equation}
	
	\emph{Proof:} For fixed $N$ and $\gamma$, every feature available at order $\mathcal{B}$ remains available at order $\mathcal{B}+1$. Similarly, increasing the tap length from $N$ to $N+1$ retains all existing features and adds features associated with $x(n-N)$. Hence, the corresponding approximation spaces satisfy
	\begin{equation}
		\mathcal{T}_{\mathcal{B}\text{,}\,N\text{,}\,\gamma}\subseteq\mathcal{T}_{\mathcal{B}+1\text{,}\,N\text{,}\,\gamma} \quad \text{and} \quad \mathcal{T}_{\mathcal{B}\text{,}\,N\text{,}\,\gamma}\subseteq\mathcal{T}_{\mathcal{B}\text{,}\,N+1\text{,}\,\gamma} \text{.}
		\label{907}
	\end{equation}
	From (\ref{907}), minimization over the larger subspace cannot increase the projection error, which proves (\ref{902}) and (\ref{903}). Let $\mathnormal{F}_\mathcal{B}k_{j\text{,}\,\gamma}$ denote the Fourier projection of $k_{j\text{,}\,\gamma}$ onto the Fourier subspace of order $\mathcal{B}$, which yields the GTFLN approximant
	\begin{equation}
		g_{\mathcal{B}\text{,}\,N\text{,}\,\gamma}(n) = c + \sum_{j=0}^{N-1}\left[\theta_j + \delta_j \mathit{X}_j+ e^{-\gamma \mathit{X}_j^2}\mathnormal{F}_\mathcal{B}k_{j\text{,}\,\gamma}\left(\mathit{X}_j\right)\right]\text{.}
		\label{908}
	\end{equation} 
	The function in (\ref{908}) belongs to the finite GTFLN space because it contains only a constant term, the linear inputs, and the GF-based TBFs. Using Parseval's identity yields
	\begin{equation}
		\left\|k_{j\text{,}\,\gamma} - \mathnormal{F}_\mathcal{B}k_{j\text{,}\,\gamma}\right\|^2_{L^2(-1\text{,}\,1)}\leq  \frac{1}{\left[\pi\left(\mathcal{B}+1\right)\right]^{2s}}\Vert k_{j\text{,}\,\gamma}^{(s)}\Vert^2_{L^2(-1\text{,}\,1)}\text{,}
		\label{909}
	\end{equation} 
	Combining $0<e^{-\gamma t^2}\leq1$ and $q_j(t)\leq \mathcal{M}_j$ with (\ref{909}) yields 
	\begin{equation}
		\left\|e^{-\gamma\mathit{X}_j^2}\left[k_{j\text{,}\,\gamma} - \mathnormal{F}_\mathcal{B}k_{j\text{,}\,\gamma}\right](\mathit{X}_j)\right\|_{L^2(P)}
		\leq 
		\frac{\sqrt{\mathcal{M}_j}\Vert k_{j\text{,}\,\gamma}^{(s)}\Vert_{L^2(-1\text{,}\,1)}}{\left[\pi\left(\mathcal{B}+1\right)\right]^s}\text{,}
		\label{910}
	\end{equation}
	Applying the triangle inequality to the approximation residual associated with (\ref{908}) yields (\ref{906}). Since $\varepsilon_{\rm app}$ is the infimum over all admissible GTFLN approximants, it is no greater than the error of the approximant in (\ref{908}). Squaring (\ref{906}) gives the corresponding upper bound on $\varepsilon_{\rm app}^2(\mathcal B\text{,}\,N\text{,}\,\gamma)$. 
	
	The result in (\ref{903}) is distribution-free, but it does not imply the target-independent convergence rate with respect to $N$. For an additive system, an omitted-memory residual satisfying $\left\|\alpha_N\right\|_{L^2(P)}\leq C\lambda^N$, where $C>0$ and $0<\lambda<1$, decreases at least geometrically. For a finite-memory additive system, this residual vanishes once $N$ reaches the true memory length. In contrast, the nonadditive cross-tap component may remain nonzero even as $N$ and $\mathcal{B}$ increase because the GTFLN dictionary is additive across taps. Eliminating this residual requires explicit cross-tap features, at the cost of increased model order and computational complexity. 
	
	The spaces associated with different values of $\gamma$ are not nested. Thus, the best approximation error is not generally monotone in $\gamma$. Assuming that $R_{\mathcal{B}\text{,}\,N\text{,}\,\gamma}$ is nonsingular, differentiation of the minimized MSE with respect to $\gamma$, together with the Wiener orthogonality condition, yields
	\begin{equation}
		\frac{\partial \varepsilon_{\rm app}^2\left(\mathcal{B}\text{,}\,N\text{,}\,\gamma\right)}{\partial \gamma} = -2\mathit{E}\left\{e_{o\text{,}\,\gamma}(n)\bm\omega_{o\text{,}\,\gamma}^\mathrm{T}\frac{\partial \mathcal{A}_{\mathcal{B}\text{,}\,N\text{,}\,\gamma}(n)}{\partial \gamma}\right\}\text{,}
		\label{911}
	\end{equation}
	where $\bm \omega_{o\text{,}\,\gamma}$ denotes the Wiener solution for the fixed $\gamma$, $e_{o\text{,}\,\gamma}(n) = \breve{d}(n)-\bm\omega_{o\text{,}\,\gamma}^\mathrm{T}\mathcal{A}_{\mathcal{B}\text{,}\,N\text{,}\,\gamma}(n)$, $\partial \mathcal{G}_{i,j}(n)/\partial\gamma = -\mathit{X}_j^2\mathcal{G}_{i,j}(n)$, and $\partial \mathcal{Q}_{i,j}(n)/\partial\gamma = -\mathit{X}_j^2\mathcal{Q}_{i,j}(n)$. The sign of (\ref{911}) depends on the system output mapping and input statistics. A small $\gamma$ provides a broad harmonic representation, whereas a large $\gamma$ increases localization but suppresses the features away from the origin. Hence, no target-independent optimal value of $\gamma$ exists. (\ref{902}) and (\ref{903}) concern the best approximation error rather than the LMS steady-state error. The expected squared norm of the feature vector satisfies $\operatorname{Tr}[\bm R_{\mathcal{B}\text{,}\,N\text{,}\,\gamma}]= \mathit{E}\{\left\|\mathcal{A}_{\mathcal{B}\text{,}\,N\text{,}\,\gamma}(n)\right\|^2\} = 1+\sum_{j=0}^{N-1}\{\mathit{E}\{\mathit{X}_j^2\}+ \mathcal{B}\mathit{E}\{e^{-2\gamma \mathit{X}_j^2}\}\}
	$, where $\sin^2\left[i\pi x(n)\right]+\cos^2\left[i\pi x(n)\right]=1$ is utilized and $\operatorname{Tr}(\cdot)$ denotes the trace of the matrix. Consequently, increasing $\mathcal{B}$ and $N$ can increase the LMS misadjustment at the fixed step size even though the best approximation error in (\ref{901}) is nonincreasing.
\subsection{Mean Square Behavior}
This subsection presents a theoretical EMSE analysis of the GTFLN filter. For analytical tractability, the following assumptions are introduced \cite{sayed2003fundamentals,patel2020convergence,1179757}.

\textit{Assumption I:} The input signal $x(n)$ and noise $v(n)$ are zero-mean, and mutually uncorrelated Gaussian random variables with unit variance and $\sigma_v^2$, respectively.

\textit{Assumption II:} The adaptive filter is sufficiently long and as such, the autocorrelation matrix of the expanded input vector $\mathcal{\bm A}(n)$ is uncorrelated with the square of the \textit{a priori} error $\zeta(n)$.

\textit{Assumption III:} For analytical tractability, the independence approximation is adopted that $\mathcal{A}(n)$ is statistically independent of  $\bm\omega(n)$, and the observation noise $v(n)$ is independent of $x(n)$, $\mathcal{A}(n)$, and $\bm\omega(n)$. Moreover,  $x(n)$ and $v(n)$ are each assumed to be independently and identically distributed (i.i.d.).

Note that Assumptions I-III are commonly used in the theoretical analysis of adaptive filters to simplify the derivation. 

For the EMSE analysis, \textit{a priori} error $\zeta(n)$ is introduced as
\begin{equation}
	\zeta(n) = \breve{d}(n) - y(n)= \bm\omega_o^\mathrm{T}\mathcal{\bm A}_o(n) - \bm\omega^\mathrm{T}(n)\mathcal{\bm A}(n)\text{.}
	\label{041}
\end{equation}
Based on (\ref{007}) and (\ref{041}) and Assumption I, the EMSE as a performance metric is defined as
\begin{equation}
	\mathcal{J}_{ex}(n) = \mathit{E}\left\{e^2(n)\right\} - \mathit{E}\left\{\left[e(n) - \zeta(n)\right]^2\right\}\text{.}
	\label{042}
\end{equation}
The weight deviation vector is defined as 
$	\widetilde{\bm \omega}(n) = \bm \omega_o - \bm\omega(n)
$ \cite{sayed2003fundamentals}. The weight vector of the GTFLN filter is updated using an LMS-type algorithm.  Subtracting $\bm \omega_o$ from both sides of (\ref{009}) and taking the $L_2$-norm yield
\begin{equation}
	\begin{aligned}
		&\left\|\widetilde{\bm\omega}(n+1)\right\|^2\\ =&\left\|\widetilde{\bm\omega}(n)\right\|^2 - 2\mu e(n)\widetilde{\bm \omega}^\mathrm{T}(n)\mathcal{\bm A}(n)+\mu^2e^2(n)\left\|\mathcal{\bm A}(n)\right\|^2\text{.}
		\label{043}
	\end{aligned}
\end{equation}
Based on (\ref{012}), the error signal $e(n)$ can be re-expressed as
\begin{equation}
	e(n) =\varrho_1(n) + \varrho_2(n) + v(n)\text{,}
	\label{045}
\end{equation}
where $\varrho_1(n) = \widetilde{\bm \omega}^\mathrm{T}(n)\mathcal{\bm A}_o(n)$
and $\varrho_2(n) = \bm\omega^\mathrm{T}(n)\bm\Gamma(n)$. Using (\ref{045}) and $\bm\Gamma(n) = \mathcal{\bm A}_o(n) - \mathcal{\bm A}(n)$, $\widetilde{\bm \omega}^\mathrm{T}(n)\mathcal{\bm A}(n)$ in (\ref{043}) can be calculated as 
\begin{equation}
	\widetilde{\bm \omega}^\mathrm{T}(n)\mathcal{\bm A}(n) 
	= \varrho_1(n) + \varrho_2(n) - \bm\omega_o^\mathrm{T}\bm\Gamma(n)\text{.}
	\label{046}
\end{equation}
Substituting (\ref{045}) and (\ref{046}) into the second and third terms of the right side of (\ref{043}) yields, respectively,
\begin{equation}
	\begin{aligned}
		\varpi_1(n) = &\;e(n)\widetilde{\bm \omega}^\mathrm{T}(n)\mathcal{\bm A}(n)\\
		=&\;\varrho_1^2(n) + \varrho_2^2(n) + 2\varrho_1(n)\varrho_2(n)+ v(n)\varrho(n) \\
		&- \varrho(n)\bm\omega_o^\mathrm{T}\bm\Gamma(n) - v(n)\bm\omega_o^\mathrm{T}\bm\Gamma(n)
		\label{047}
	\end{aligned}
\end{equation}
and
\begin{equation}
	\begin{aligned}
		\varpi_2(n) = &\;e^2(n)\left\|\mathcal{\bm A}(n)\right\|^2\\	
		= &\left[\varrho_1^2(n) + \varrho_2^2(n) + v^2(n)\right]\left\|\mathcal{\bm A}(n)\right\|^2\\
		&+\left[2\varrho(n)v(n) + 2\varrho_1(n)\varrho_2(n)\right]\left\|\mathcal{\bm A}(n)\right\|^2\text{,}
		\label{048}
	\end{aligned}
\end{equation}
where $\varrho(n) = \varrho_1(n) + \varrho_2(n)$. At steady state, assume that $\lim\limits_{n\to\infty}\mathit{E}\left\{\left\|\widetilde{\bm\omega}(n+1)\right\|^2\right\}= \lim\limits_{n\to\infty}\mathit{E}\left\{\left\|\widetilde{\bm\omega}(n)\right\|^2\right\}$. By substituting (\ref{047}) and (\ref{048}) into (\ref{043}) and taking expectation on both sides under Assumptions I-III, (\ref{043}) can be simplified as
\begin{equation}
	\begin{aligned}
		0=&\left\{2-\mu\mathit{E}\left\{\left\|\mathcal{\bm A}(n)\right\|^2\right\}\right\}\mathit{E}\left\{\varrho^2_1(n) + \varrho^2_2(n)\right\}\\
		&-2\mathit{E}\left\{\varrho_2(n)\bm\omega_o^\mathrm{T}\bm\Gamma(n)\right\}-\mu\sigma_v^2\mathit{E}\left\{\left\|\mathcal{\bm A}(n)\right\|^2\right\}\text{.}
	\end{aligned}
	\label{044}
\end{equation}
The following simplifications are utilized to obtain (\ref{044}) at steady state
\begin{equation}
	\begin{aligned}
		&\lim\limits_{n\to\infty}\mathit{E}\left\{\varrho_1(n)\bm\omega_o^\mathrm{T}\bm\Gamma(n)\right\}\\
		= &\lim\limits_{n\to\infty}\left\{\bm\omega_o - \mathit{E}\left\{\bm \omega(n)\right\}\right\}^\mathrm{T}\mathit{E}\left\{\mathcal{\bm A}_o(n)\bm\Gamma^\mathrm{T}(n)\right\}\bm\omega_o\approx 0
	\end{aligned}
	\label{049}
\end{equation}
and
\begin{equation}
	\begin{aligned}
		&\lim\limits_{n\to\infty}\mathit{E}\left\{\varrho_1(n)\varrho_2(n)\right\}\\
		=& \lim\limits_{n\to\infty}\left\{\bm\omega_o - \mathit{E}\left\{\bm \omega(n)\right\}\right\}^\mathrm{T}\mathit{E}\left\{\mathcal{\bm A}_o(n)\varrho_2(n)\right\}\approx 0
		\text{.}
		\label{050}
	\end{aligned}
\end{equation}

From (\ref{042}), we get $\mathcal{J}_{ex}(n)= \mathit{E}\left\{\varrho_1^2(n)\right\} + \mathit{E}\left\{\varrho_2^2(n)\right\}$. Hence, the EMSE performance of the GTFLN filter can be calculated at steady state
\begin{equation}
	\begin{aligned}
		\mathcal{J}_{ex}&=\lim\limits_{n\to\infty}\mathcal{J}_{ex}(n)=\lim\limits_{n\to\infty}\left\{\mathit{E}\left\{\varrho_1^2(n)\right\} + \mathit{E}\left\{\varrho_2^2(n)\right\}\right\}\\
		&=\frac{2\operatorname{Tr}\left[\bm \Xi(n)\right] + \mu\sigma_v^2\operatorname{Tr}\left(\bm \Upsilon_1\right)}{2-\mu \operatorname{Tr}\left(\bm \Upsilon_1\right)}\text{.}
		\label{051}
	\end{aligned}
\end{equation}
Since both expectation and transposition are linear operations, $\operatorname{Tr}\left[\bm \Xi(n)\right]$ and $\operatorname{Tr}\left(\bm \Upsilon_1\right)$ are calculated by 
\begin{equation}
	\begin{aligned}
		\operatorname{Tr}[\bm \Xi(n)] &= \mathit{E}\left\{\varrho_2(n)\bm\omega_o^\mathrm{T}\bm\Gamma(n)\right\}\\
		&=\operatorname{Tr}\left\{\mathit{E}\left\{\bm\Gamma(n)\bm\Gamma^\mathrm{T}(n)\right\}\mathit{E}\left\{\bm\omega_o\bm\omega^\mathrm{T}(n)\right\}\right\}
		\label{053}
	\end{aligned}
\end{equation}
and
\begin{equation}
	\begin{aligned}
		\operatorname{Tr}\left(\bm \Upsilon_1\right) = \mathit{E}\left\{\left\|\mathcal{\bm A}(n)\right\|^2\right\} =\operatorname{Tr}\left\{\mathit{E}\left\{\mathcal{ \bm A}(n)\mathcal{ \bm A}^\mathrm{T}(n)\right\}\right\}\text{,}
	\end{aligned}
\end{equation}
where $\bm \Xi(n) = \mathit{E}\left\{\bm\Gamma(n)\bm\Gamma^\mathrm{T}(n)\right\}\mathit{E}\left\{\bm\omega_o\bm\omega^\mathrm{T}(n)\right\}$.

\section{Simulation Results}
\label{simulations}
This section is divided into four parts to evaluate the proposed nonlinear filters. In the first part, the effectiveness of the theoretical steady-state EMSE analysis is verified through two different nonlinear NSI scenarios. In the second part, the efficacy of the GTFLN and OGTFLN filters is validated by four different NSI scenarios. In the third part, the practical applicability of the GTFLN filter is examined for the speech-based NAEC. In the fourth part, the noise reduction performance of the FgLMS algorithm is validated through three different NANC scenarios. Note that both the first and second parts are conducted in NSI settings but address different validation objectives.
All simulated results are obtained by averaging over 100 independent trials.


\subsection{Validation of Theoretical Analysis in Nonlinear System Identification}
	This subsection focuses on validating the theoretical steady-state EMSE analysis derived in (\ref{051}) by comparing the analytical findings with the simulated steady-state results.

\begin{figure}[htbp]
	\centering
	\begin{minipage}{0.492\linewidth}
		\centering
		\subfigure[]{\includegraphics[width=1.05\linewidth]{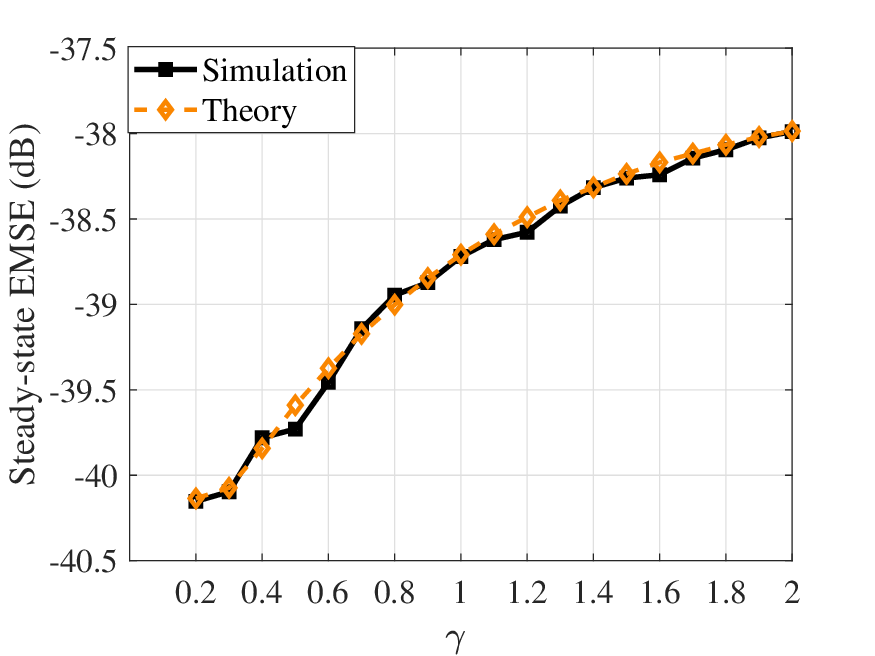}}
	\end{minipage}
	\begin{minipage}{0.492\linewidth}
		\centering
		\subfigure[]{\includegraphics[width=1.05\linewidth]{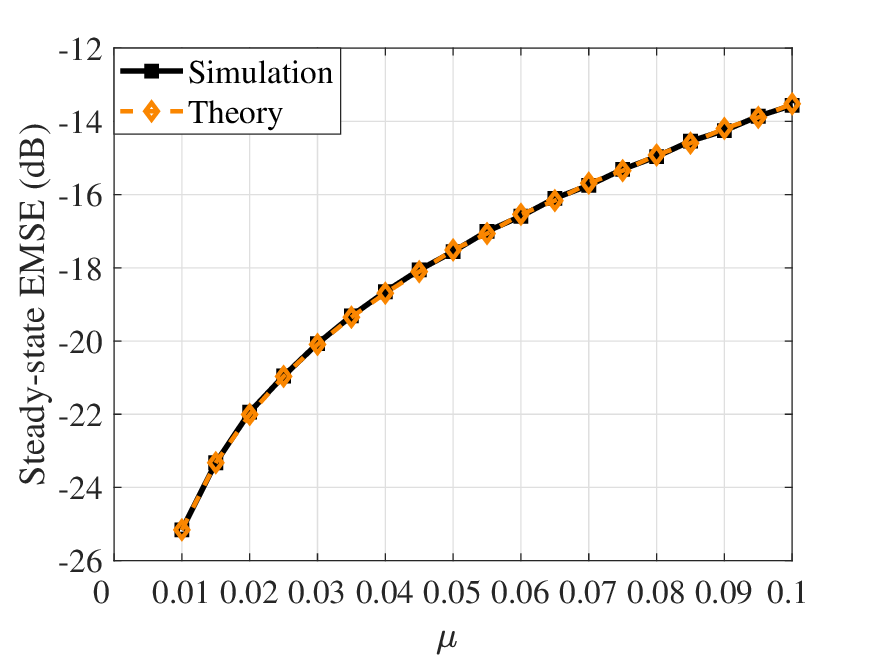}}
	\end{minipage}
	\caption{Comparison of steady-state EMSEs between simulation and theory. (a) scaling parameter $\gamma$ with $\mu=5\times10^{-3}$ and SNR $= 20$ dB. (b) Step size $\mu$ with $\gamma = 0.8$ and SNR $=10$ dB.}
	\label{theory_1}
\end{figure}

\textit{Experiment 1:} The validity of the theoretical steady-state EMSE analysis for the GTFLN filter updated by the GTFLMS algorithm is verified in the context of NSI. The unknown nonlinear system is described by (\ref{004}) with $\mathcal{B}=2$, $N=2$, and $\gamma=0.8$.
The input signal $x(n)$ is zero-mean WGN with unit variance. The weight vector $\bm\omega_o$ of the unknown nonlinear system is considered as $\bm\omega_o=[0.8\text{,}-0.6\text{,}\,0.3\text{,}-0.1\text{,}\,0.2\text{,}-0.7\text{,}\,0.4\text{,}\,0.5\text{,}-0.9\text{,}\,0.3\text{,}\,0.6]^\mathrm{T}$. To validate the theoretical EMSE analysis of the GTFLN filter, simulations are conducted under additive Gaussian noise across two examples. In the first example, the scaling parameter is varied over $\gamma=[0.2\text{,}\;2]$ with the step size $\mu=5\times10^{-3}$ and the signal-to-noise ratio (SNR) $=20$ dB, where SNR, dB $= 10\log_{10}\left\{\frac{P_x}{\sigma_v^2}\right\}$ and $P_x$ denotes the average power of $x(n)$. In the second example, the step size is varied within $\mu = \left[0.01\text{,}\;0.1\right]$ with $\gamma=0.8$ and SNR $=10$ dB. From Figs. \ref{theory_1}(a) and (b), the simulated EMSEs exhibit excellent agreement with the theoretical results.

\textit{Experiment 2:} To further validate the effectiveness of the theoretical analysis, the nonlinear system with an asymmetric loudspeaker distortion is considered. The nonlinear relation between the input and output is expressed as \cite{patel2016design}
\begin{equation}
	\breve{d}(n) = \beta\left\{\frac{1}{1 + \exp\left[-\eta r(n)\right]} - \frac{1}{2}\right\}\text{,}
	\label{061}
\end{equation}
where $\beta$ represents the system gain and is chosen as $\beta=2$, $\eta$ denotes the slope of a sigmoid function with $\eta=1/2$ for $r(n)<0$ and $\eta=4$ for $r(n)\geq0$, respectively, and $r(n) = \frac{3}{2}x(n)-\frac{3}{10}x^2(n)$. The desired output $d(n)$ of the NSI is obtained by convolving the output $\breve{d}(n)$ of the memoryless nonlinear system with the 128-tap acoustic impulse response sampled at a rate of $f_s = 8$ kHz between the loudspeaker and microphone. The impulse response of the acoustic path is adopted from \cite{7892916}.
The input signal $x(n)$ is the zero-mean WGN with $x(n)\sim\mathcal{N}(0\text{,}\;1)$ and SNRs of $\left[0 \;\text{dB}\text{,}\; 10 \;\text{dB}\text{,}\;20 \;\text{dB}\right]$ are considered.
\begin{figure}[!htbp]
	\centering
	\includegraphics[width=1.025\columnwidth]{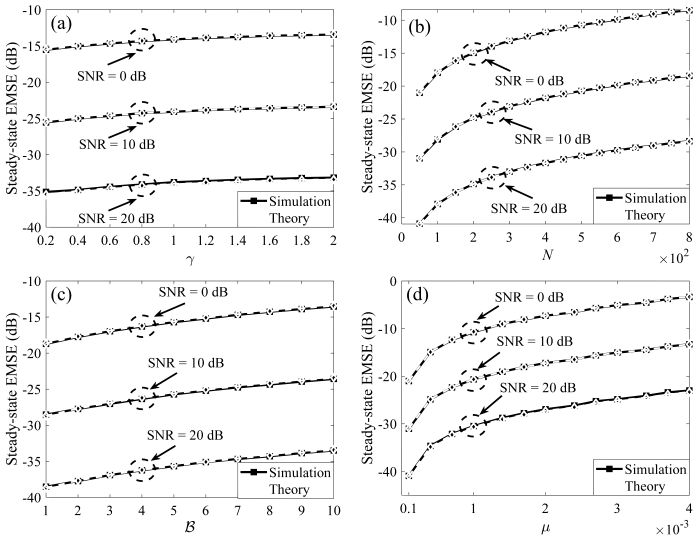}
	\caption{Comparison of steady-state EMSEs between theory and simulation under different SNRs. (a) scaling parameter $\gamma$ with $\mu = 4\times10^{-4}$. (b) Tap-length $N$ with $\mu = 2\times10^{-4}$ and $\gamma=0.5$. (c) Expansion order $\mathcal{B}$ with $\mu = 2\times10^{-4}$ and $\gamma=0.5$. (d) Step size $\mu$ with $\gamma=0.5$ and $\mathcal{B}=1$.}
	\label{theory_2_2}
\end{figure}

To comprehensively validate the theoretical steady-state EMSE analysis derived in (\ref{051}) for the GTFLN filter, four simulations are conducted, including the scaling parameter $\gamma=[0.2\text{,}\;2]$ with $\mu=4\times10^{-4}$, $\mathcal{B} = 1$, and $N = 100$, the tap-length $N=\left[50\text{,}\;800\right]$ with $\gamma=0.5$, $\mu=2\times10^{-4}$, $\mathcal{B}=1$, expansion order $\mathcal{B}=\left[1\text{,}\;10\right]$ with $\mu=2\times10^{-4}$, $\gamma=0.5$, and $N = 100$, and step size $\mu=\left[10^{-4}\text{,}\;4\times10^{-3}\right]$ with $\gamma = 0.5$, $\mathcal{B} = 1$, and $N = 100$. As depicted in Figs. \ref{theory_2_2}(a)–(d), the theoretical EMSEs exhibit excellent agreement with the simulated results across four examples, confirming the analytical accuracy. From Fig. \ref{theory_2_2}(a) and (c), a smaller $\gamma$ yields a lower EMSE, highlighting the sensitivity of the GTFLN filter to $\gamma$, and an increase in the expansion order $\mathcal{B}$ yields a higher steady-state error. However, the slight theoretical deviation observed at large step sizes in Fig. \ref{theory_2_2}(d) can be attributed to the simplifying assumptions in deriving (\ref{051}).

\subsection{Nonlinear System Identification}
	In this subsection, the performance of the GTFLN and OGTFLN filters is evaluated in terms of convergence and misadjustment in four different nonlinear system scenarios for NSI. In all experiments, the initial weight vector $\bm \omega(n)$ is set to a zero vector. The MSE is used as the performance metric.

\emph{Experiment 1:} In the first example, the input-output relation of the nonlinear system is described by the asymmetric sigmoid loudspeaker distortion model in (\ref{061}). 

\emph{Experiment 2:} In the second example, the input-output relation of the nonlinear system is introduced as the memoryless loudspeaker distortion with a soft clipping function \cite{patel2016design}
\begin{equation}
	\breve{d}(n) = \left\{
	\begin{aligned}
		&2x(n)/\left(3\rho\right) & & 0 \leq \left|x(n)\right| <\rho \\
		&\text{sign}\left[x(n)\right]\frac{3-\left[2-\left|x(n)\right|/\rho\right]^2}{3} & & \rho \leq \left|x(n)\right| < 2\rho \\
		&\text{sign}\left[x(n)\right] & & 2\rho \leq \left|x(n)\right| \leq 1\text{,}
	\end{aligned}
	\right.
	\label{065}
\end{equation}
where $0<\rho\leq0.5$ represents the threshold and $\text{sign}\left[\cdot\right]$ stands for the sign function. The nonlinear system model can characterize the nonlinear nature of the loudspeaker. 

\emph{Experiment 3:} In the third example, the input-output relation of the nonlinear system is chosen as \cite{patel2016design}
\begin{equation}
	\begin{aligned}
		\breve{d}(n) = \exp\left[0.5x(n)\right]\left\{\!\right.&\sin\left[\pi x(n)\right] + 0.3\sin\left[3\pi x(n)\right] \\
		&+ 0.1\sin\left[5\pi x(n)\right]\!\left.\right\}\text{.}
	\end{aligned}
\end{equation}

\emph{Experiment 4:} In the fourth example, the input-output relationship of the nonlinear system is given by \cite{patel2016design}
\begin{equation}
	\breve{d}(n)=0.6\sin^3\left[\pi x(n)\right] - \frac{2}{x^3(n) + 2} - 0.1\cos\left[4\pi x(n-4)\right] + \frac{5}{4}\text{.}
\end{equation}

\begin{figure}[htbp]
	\centering
	\begin{minipage}{0.49\linewidth}
		\centering
		\subfigure[]{\includegraphics[width=1.05\linewidth]{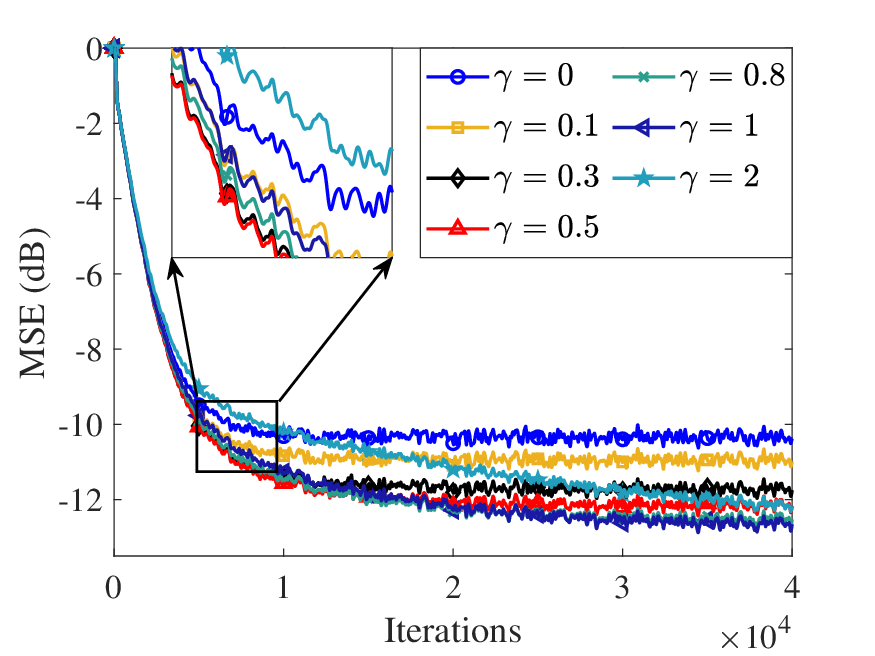}}
	\end{minipage}
	\begin{minipage}{0.49\linewidth}
		\centering
		\subfigure[]{\includegraphics[width=1.05\linewidth]{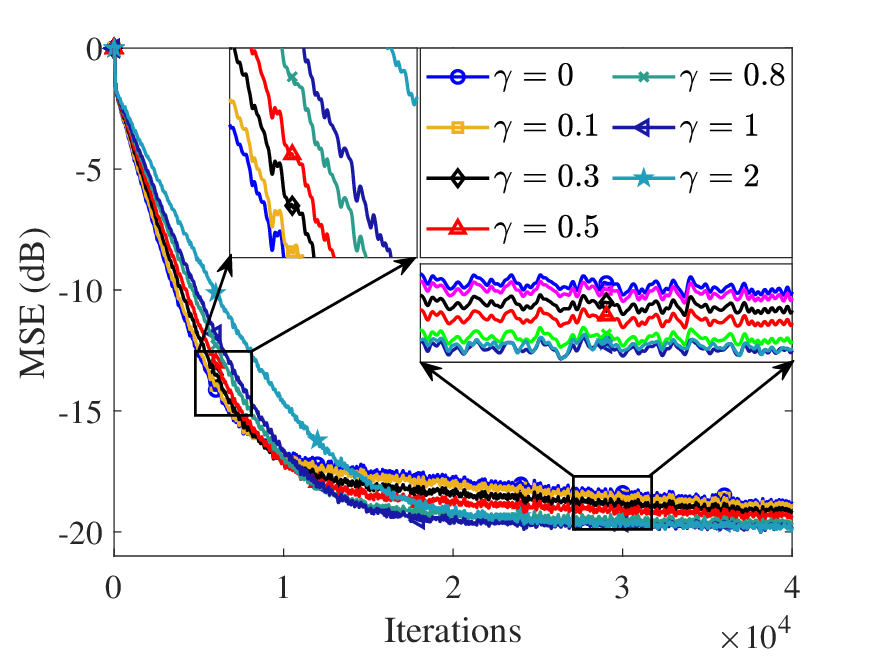}}
	\end{minipage}
	\caption{Comparison of the GTFLN filter with different scaling parameter $\gamma$. (a) Gaussian input. (b) Uniformly distributed input.}
	\label{gamma_para}
\end{figure}

Before carrying out performance comparison of the GTFLN and OGTFLN filters over the benchmarks for the NSI, the effects with respect to $\gamma$ for the GTFLN filter should be determined, where a Hammerstein system consisting of the soft clipping function described in (\ref{065}) followed by a 4-tap finite impulse response (FIR) filter is considered. Note that when $\gamma=0$, the GTFLN filter is reduced to the TFLN filter. Let $\alpha(n)$ denote the output of the soft clipping function with $\rho=0.3$. The output of the unknown system is given by $\breve{d}(n) = 0.8\alpha(n)+0.5\alpha(n-1)-0.3\alpha(n-2)+0.1\alpha(n-3)$. The input is chosen as a zero-mean unit-variance Gaussian input and a uniformly distributed input with range $[-1\text{,}\,1]$. In both cases, the observation noise $v(n)$ is zero-mean WGN with variance $\sigma_v^2=0.01$.  The Gaussian and uniform inputs yield SNRs of $20$ dB and $15.23$ dB, respectively. Fig. \ref{gamma_para} plots the comparison of the GTFLN filter with different $\gamma=\left[0\text{,}\,0.1\text{,}\,0.3\text{,}\,0.5\text{,}\,0.8\text{,}\,1\text{,}\,2\right]$ with $\mu=4\times10^{-4}$, $\mathcal{B}=2$, and $N=10$. It is observed that a small $\gamma$ yields fast convergence, and a large $\gamma$ yields a small steady-state error. The GTFLN filter with $\gamma=0.5$ is selected as the compromise that provides fast convergence and a comparable steady-state error. In the subsequent NSI simulations, $\gamma$ is set to $0.5$ for the GTFLN filter.

To maintain a focused comparison within the finite-dimensional LIP framework, the SOV \cite{1453790}, TFLN \cite{das2004active}, GeTFLN \cite{sicuranza2011generalized}, and AETFLN \cite{patel2016design} filters using LMS-type adaptation are employed as the benchmarks. The SOV filter represents a polynomial expansion, the TFLN filter is the parent structure of the GTFLN filter, the GeTFLN filter represents a cross-term extension, and the AETFLN filter is the closest envelope-based counterpart. For Experiments 1-4, the expansion structures of the nonlinear filters are listed in Table \ref{Table_NSI}. For a fair comparison, the parameter settings of the GTFLN and OGTFLN filters with the benchmark filters are manually chosen to achieve comparable initial convergence. Particularly, the number of cross-terms for the GeTFLN filter is selected as $P=2$.
\begin{table}[htp]
	\small
	\centering
	\scriptsize
	\caption{Comparison of Computational Complexities and Memory Requirements For Nonlinear Filters Per Iteration under Experiments 1-4}
	\label{Table_NSI}
	\renewcommand{\arraystretch}{1.7}
	\doublerulesep=0.6pt
	\begin{tabular}{cccccccc}
		\hline\hline
		\textbf{filters} &$N$ &$\mathcal{B}$ &$L$ &\textbf{$\times/\div$} &\textbf{$+/-$} &{$\sin(\cdot)/\cos(\cdot)$} & Memory \\ \hline
		\textbf{SOV}           &$12$ &$0$   &$90$  &$259$ &$180$ &$0$ &$196$ \\ \hline
		\textbf{TFLN}            &$15$ &$2$  &$75$ &$181$ &$150$ &$60$ &$154$ \\ \hline
		\textbf{GeTFLN}				    &$10$ &$1$  &$64$ &$173$ &$128$ &$20$ &$240$ \\ \hline
		\textbf{AETFLN}          &$10$ &$2$  &$51$ &$277$ &$153$ &$40$ &$180$ \\ \hline
		\textbf{GTFLN}           &$10$ &$2$  &$51$ &$184$ &$102$ &$40$ &$117$ \\ \hline
		\textbf{OGTFLN}          &$10$ &$2$  &$51$ &$401$ &$255$ &$40$ &$173$ \\ \hline\hline
	\end{tabular}
\end{table}

\begin{figure}[htbp]
	\centering
	\begin{minipage}{0.492\linewidth}
		\centering
		\subfigure[]{\includegraphics[width=1.05\linewidth]{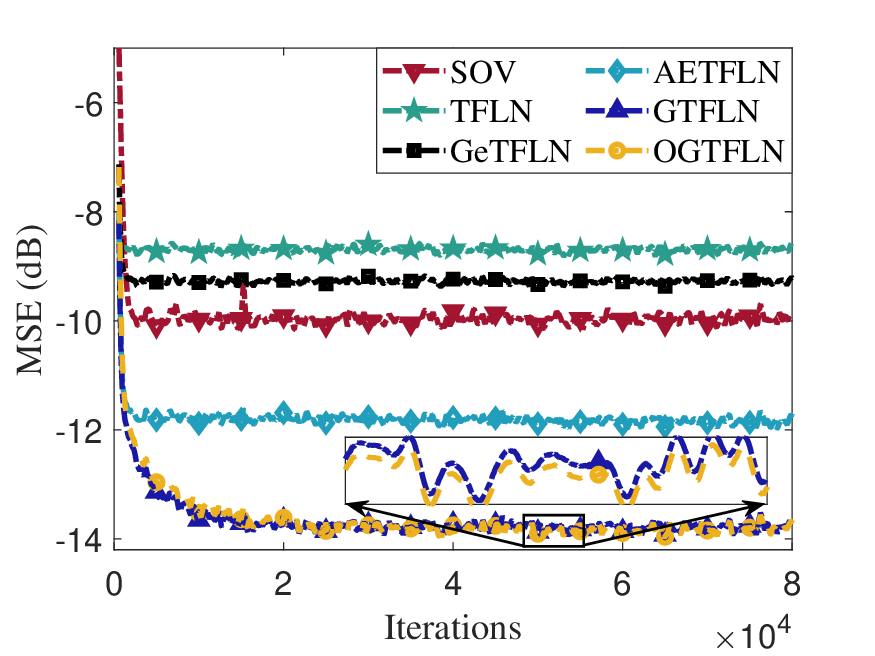}}
	\end{minipage}
	\begin{minipage}{0.492\linewidth}
		\centering
		\subfigure[]{\includegraphics[width=1.05\linewidth]{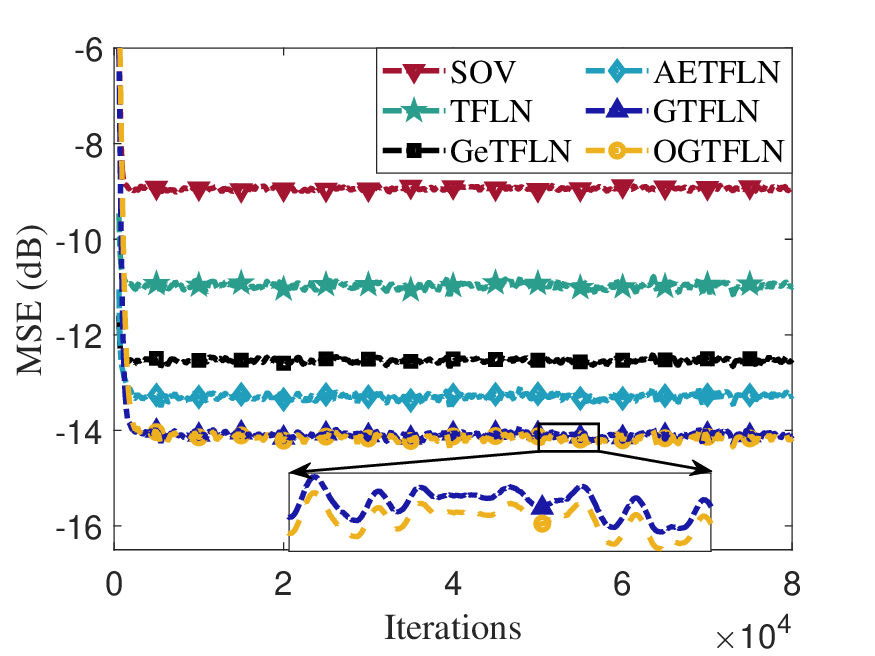}}
	\end{minipage}
	
	\begin{minipage}{0.492\linewidth}
		\centering
		\subfigure[]{\includegraphics[width=1.05\linewidth]{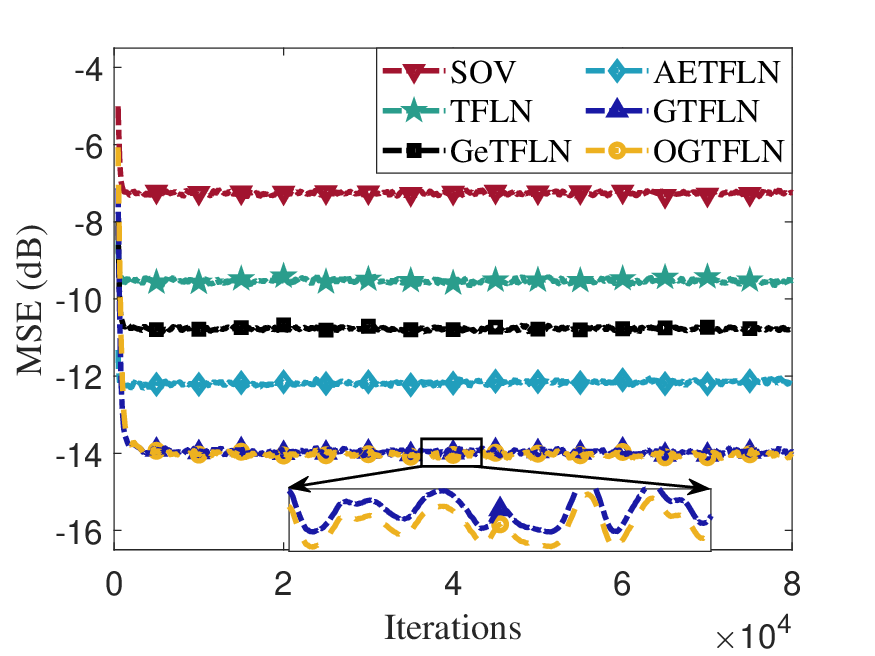}}
	\end{minipage}
	\begin{minipage}{0.492\linewidth}
		\centering
		\subfigure[]{\includegraphics[width=1.05\linewidth]{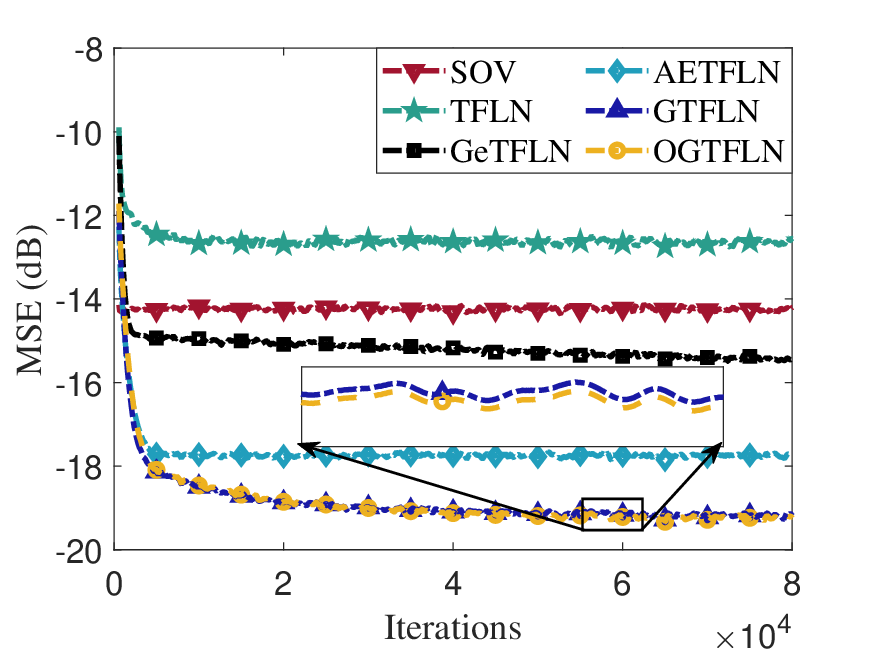}}
	\end{minipage}
	\caption{MSEs of the SOV, TFLN, GeTFLN, AETFLN, GTFLN, and OGTFLN filters. (a) Experiment 1. (b) Experiment 2. (c) Experiment 3. (d) Experiment 4.}
	\label{NSI_EX}
\end{figure}

In Experiment 1, the MSE performance of the SOV \cite{1453790}, TFLN \cite{das2004active}, GeTFLN \cite{sicuranza2011generalized}, AETFLN \cite{patel2016design}, GTFLN, and OGTFLN filters is compared in Fig. \ref{NSI_EX}(a). The input and observation noise $v(n)$ are the zero-mean WGN with variances 2 and $0.001$, respectively, corresponding to an SNR of $33.01$ dB. The parameters are set to $\mu=8\times10^{-4}$ for the SOV, $\mu=6\times10^{-3}$ for the TFLN, $\mu=2\times10^{-3}$ for the GeTFLN, $\mu=2\times10^{-3}$ and $\mu_a = 10^{-4}$ for the AETFLN, $\gamma=0.5$ and $\mu=2\times10^{-3}$ for the GTFLN, and $\mu=1.5\times10^{-3}$ for the OGTFLN with $\gamma_o(n)$ calculated by (\ref{028}). As can be seen, 
the GTFLN and OGTFLN filters outperform the benchmarks, achieving the lowest steady-state misadjustments. 

In Experiments 2 and 3, the input is uniformly distributed within $\left[-1\text{,}\;1\right]$, and parameters are adjusted to achieve comparable convergence. For Experiment 2, the step size is $\mu=10^{-2}$, $\mu=3\times10^{-2}$, $\mu=2.4\times10^{-2}$, and $\mu=1.6\times10^{-2}$ and $\mu_a=10^{-3}$ for the SOV, TFLN, GeTFLN, and AETFLN, respectively, while $\mu=6\times10^{-3}$ and $\gamma=0.5$ for the GTFLN, and $\mu=4\times10^{-3}$ for the OGTFLN. For Experiment 3, the step size is set to $\mu=10^{-2}$, $\mu=2\times10^{-2}$, $\mu=8\times10^{-3}$, and $\mu=3\times10^{-2}$ and $\mu_a=10^{-3}$ for the SOV, TFLN, GeTFLN, and AETFLN, respectively, and $\mu=8\times10^{-3}$ and $\gamma=0.5$ for the GTFLN, and $\mu=6\times10^{-3}$ for the OGTFLN. In both experiments, the output of the unknown nonlinear system is corrupted by observation noise with variance of $10^{-3}$, corresponding to an SNR of $25.23$ dB. The MSEs of the GTFLN and OGTFLN filters are shown in Fig. \ref{NSI_EX}(b) and (c) compared to the benchmarks. It is clear that both the proposed filters demonstrate superior performance in terms of misadjustment, which is attributed to the GF-based framework that enhances modeling capability.

\begin{table}[htp]
	\small
	\centering
	\scriptsize
	\caption{Steady-State MSEs of the algorithms under Experiments 1-4 by Averaging the Last 2000 Iterations.}
	\renewcommand{\arraystretch}{1.3}
	\doublerulesep=0.6pt
	\begin{tabular}{m{1.5cm}<{\centering}|cccc}
		\hline\hline
		\multirow{1}{*}{\textbf{filters}} 
		& \multicolumn{1}{c}{Ex. \!1} & \multicolumn{1}{c}{Ex. \!2} & \multicolumn{1}{c}{Ex. \!3} &Ex. \!4 \\ \hline
		\textbf{SOV}	& \multicolumn{1}{c}{$-9.94$} & \multicolumn{1}{c}{$-8.93$} & \multicolumn{1}{c}{$-7.24$} 
		&$-14.25$ \\ 	
		\textbf{TFLN}	& \multicolumn{1}{c}{$-8.68$} & \multicolumn{1}{c}{$-10.96$} & \multicolumn{1}{c}{$-9.54$}
		&$-12.67$ \\ 
		\textbf{GeTFLN}	& \multicolumn{1}{c}{$-9.27$} & \multicolumn{1}{c}{$-12.54$} & \multicolumn{1}{c}{$-10.78$}
		&$-15.45$ \\ 
		\textbf{AETFLN} 	& \multicolumn{1}{c}{$-11.86$} & \multicolumn{1}{c}{$-13.27$} & \multicolumn{1}{c}{$-12.12$}
		&$-17.75$ \\ 
		\textbf{GTFLN}	& \multicolumn{1}{c}{$\bm{-13.70}$} & \multicolumn{1}{c}{$\bm{-14.10}$} & \multicolumn{1}{c}{$\bm{-13.96}$} 
		&$\bm{-19.19}$ \\ 
		\textbf{OGTFLN}	& \multicolumn{1}{c}{$\bm{-13.80}$} & \multicolumn{1}{c}{$\bm{-14.16}$} & \multicolumn{1}{c}{$\bm{-14.02}$}
		&$\bm{-19.25}$ 
		\\ \hline\hline
	\end{tabular}
	\label{NSI_TABLE}
\end{table}

In Experiment 4, the input is uniformly distributed over the range $\left[-0.5\text{,}\;0.5\right]$ and the observation noise is the zero-mean WGN with a variance of 0.01, corresponding to an SNR of $9.21$ dB. We set the step size $\mu=0.3$ for the SOV, $\mu=5\times10^{-2}$ for the TFLN, $\mu=2.4\times10^{-3}$ for the GeTFLN, $\mu=6\times10^{-3}$ and $\mu_a=4\times10^{-3}$ for the AETFLN, $\mu=1.2\times10^{-2}$ and $\gamma=0.5$ for the GTFLN, and $\mu=10^{-2}$ for the OGTFLN. Fig. \ref{NSI_EX}(d) plots a comparison of the GTFLN and OGTFLN filters with the benchmarks. Remarkably, the TFLN filter suffers from higher steady-state MSE than the other nonlinear filters.  To further show the efficacy of the proposed filters, the computational complexities and memory requirements, and averaged MSEs of the GTFLN and OGTFLN filters are provided in Tables \ref{Table_NSI} and \ref{NSI_TABLE}, respectively, compared with those of the SOV, TFLN, GeTFLN, and AETFLN filters under Experiments 1–4. One can see that the GTFLN filter can achieve reduced computational complexity compared to the AETFLN filter.  Compared with the memory requirements of all other nonlinear filters, the GTFLN filter has the lowest memory requirement. From Table \ref{NSI_TABLE}, the GTFLN filter can achieve improved performance compared to the benchmarks and the OGTFLN filter achieves slightly better performance at a higher computational cost, where the improvement is marginal because $\gamma$ is selected as a compromise between the convergence and steady-state error. Moreover, the condition $\gamma_o(n) - \gamma = \mathcal{O}\left(\mu\right)$ indicates that the OGTFLN filter performs a local refinement around the nominal value.
\subsection{Nonlinear Acoustic Echo Cancellation}
\begin{figure}[!htbp]
		\centering
		\begin{minipage}{0.492\linewidth}
			\centering                             
			\subfigure[]{\includegraphics[width=1.02\linewidth]{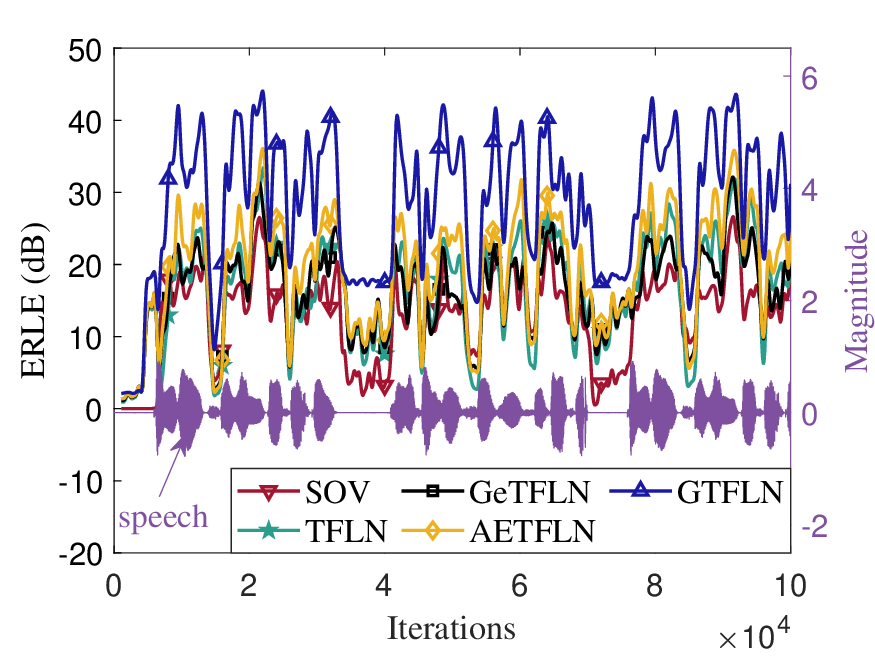}}
		\end{minipage}
		\begin{minipage}{0.492\linewidth}
			\centering
			\subfigure[]{\includegraphics[width=1.02\linewidth]{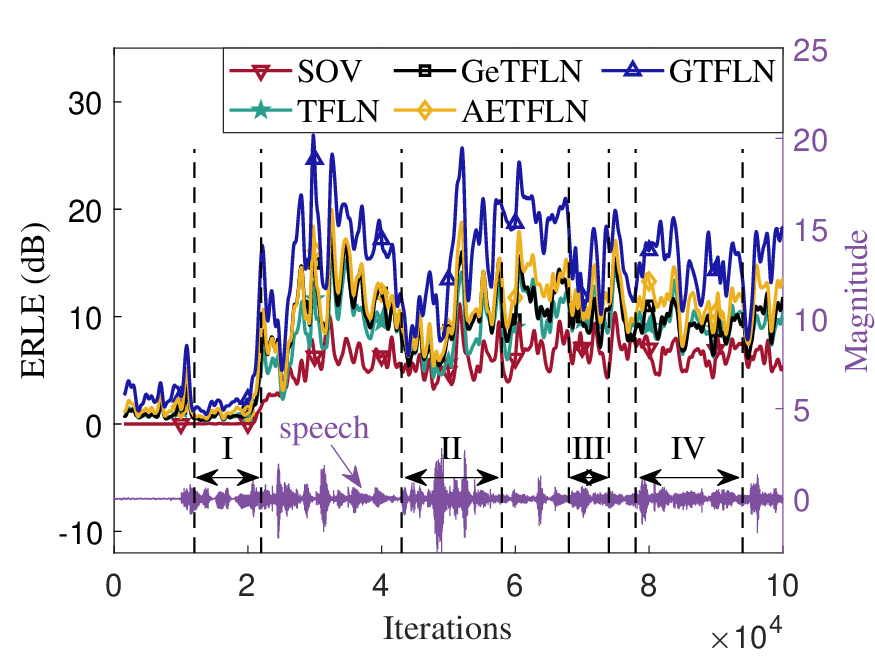}}
		\end{minipage}
		\caption{ERLE comparison of the SOV, TFLN, GeTFLN, AETFLN, and GTFLN filters for NAEC. (a) Single-talk $\xi(n)=0$. (b) Double-talk $\xi(n)\neq0$.}
		\label{NAEC}
	\end{figure}

	This subsection further evaluates the practical transferability of the GTFLN filter to NAEC. In the NAEC experiment, the nonlinear echo path is modeled as a Hammerstein system comprising a memoryless soft-clipping nonlinearity followed by a 512-tap FIR acoustic path \cite{ZHENG2026110111}. Let $\bm x(n)$ denote the far-end input signal. The nonlinear loudspeaker output is given by $u(n)=f_{\mathrm{sc}}[x(n)]$, where $f_{\mathrm{sc}}(\cdot)$ is the memoryless soft-clipping nonlinearity defined in (\ref{065}) with $\rho=0.1$. The nonlinear echo $\breve{d}(n)$ is subsequently generated by propagating $u(n)$ through the linear acoustic echo path between the loudspeaker and microphone, i.e., $\breve{d}(n)= h(n)\ast u(n)$, where $\ast$ represents the convolution operation and $h(n)$ denotes the acoustic impulse response with $T$ taps. The microphone signal is modeled as $d(n)=\breve{d}(n)+ \xi(n)+v(n)$, where $\xi(n)$ and  $v(n)$ denote the near-end input signal and background noise, respectively.  An adaptive echo canceller is designed to suppress the echo in $d(n)$, yielding the error microphone signal $e(n)$. Two experiments are considered to evaluate the GTFLN filter against the benchmarks, including the SOV, TFLN, GeTFLN, and AETFLN filters. The echo return loss enhancement (ERLE) is adopted as the performance metric and defined as \cite{6488741}
	\begin{equation}
		\text{ERLE}(n)\text{,}\; \text{dB} = 10\log_{10}\left\{\frac{\mathit{E}\{d^2(n)\}}{\mathit{E}\{e^2(n)\}}\right\}\text{.}
	\end{equation} 
	Real speech signals from the Interspeech 2021 dataset are used as the far-end inputs with a sampling rate of 8 kHz and a duration of $10$ s \cite{Jointonlinemultic2021}. The expansion structures are used with $\mathcal{B}=2$, $N=80$, and $L=401$ for the GTFLN, $\mathcal{B}=2$, $N=80$, and $L=400$ for the TFLN, $\mathcal{B}=2$, $P = 2$, $N= 35$, and $L=443$ for the GeTFLN, $\mathcal{B}=2$, $N=80$, and $L=401$ for the AETFLN, and $N=30$ and $L=495$ for the SOV. Fig. \ref{NAEC} compares the ERLEs of the GTFLN with the benchmarks. 
	
	In the single-talk scenario [Fig. \ref{NAEC}(a)], male Mandarin speech is used as the far-end signal. The observation noise has a variance of $0.001$ with an SNR of $13.05$ dB. For a fair performance comparison, the step sizes are selected as  $\mu=5\times10^{-2}$ for the SOV, $\mu=10^{-4}$ for the TFLN, $\mu=4\times10^{-4}$ for the GeTFLN, $\mu=3\times10^{-4}$ and $\mu_a = 2\times10^{-4}$ for the AETFLN, and $\mu=6\times10^{-3}$ and $\gamma=1$ for the GTFLN.  In the double-talk scenario [Fig. \ref{NAEC}(b)], male and female English speech signals are used as the far-end and near-end signals, respectively. The observation noise has a variance of $0.01$ with an SNR of $5.72$ dB.
	To mitigate the effect of double-talk, the Geigel double-talk detection (DTD) is employed to suspend the adaptation of the nonlinear filters when $\vert d(n)\vert\geq\chi\max\{\vert x(n)\vert\text{,}\,\vert x(n-1)\vert\text{,}\,\cdots\text{,}\,\vert x(n-L+1)\vert\}$ \cite{9690620}, where $\chi$ denotes the DTD threshold. In this experiment, $\chi=1$ and $L$ is set to the length of the expanded input vector for each nonlinear filter. The dashed intervals I–IV indicate the double-talk periods. To reach a comparable initial convergence, the parameter settings are selected as the SOV with $\mu=8\times10^{-4}$, the TFLN with $\mu=10^{-4}$, the GeTFLN with $\mu=2\times10^{-4}$, the AETFLN with $\mu=2\times10^{-4}$ and $\mu_a = 10^{-4}$, and the GTFLN with $\mu=8\times10^{-4}$ and $\gamma=0.5$.
	As shown in Fig. \ref{NAEC}, the GTFLN filter outperforms the benchmarks. This improvement can be attributed to the GF-based basis functions, which capture localized variations in the nonlinear input-output mapping more effectively, thereby enhancing the modeling accuracy.
\subsection{Nonlinear Active Noise Control}
In the feedforward GTFLN-based NANC system, $\bm x(n)$ denotes the reference noise sampled by the reference microphone, $d(n)$ represents the primary noise, and $y(n)$ is the controller output \cite{patel2016design}. The transfer functions of the primary and secondary paths are given by $P(z)$ and $S(z)$, respectively, with both paths modeled as FIR filters. The anti-noise signal $\widehat{y}(n)$ is generated through the secondary path $S(z)$ and has the same amplitude as $d(n)$ but is in anti-phase. To compensate for the acoustic superposition between $d(n)$ and $\widehat{y}(n)$,
the reference noise $\bm x(n)$ needs to be preprocessed using the estimated secondary path $\widehat{S}(z)$ \cite{YIN2024110887}. The mismatch between $\widehat{S}(z)$ and $S(z)$ can cause performance degradation and even divergence. Hence, the matched secondary path model is adopted to demonstrate the improved noise reduction of the FgLMS algorithm using the GTFLN filter over the benchmarks using the TFLN-based expansions for NANC systems.

The adaptive controller, implemented as the GTFLN filter, updates the weight vector $\bm \omega(n)$ using the LMS algorithm by minimizing $\mathit{E}\left\{e^2(n)\right\}$, yielding the FgLMS algorithm, where the residual noise $e(n)$ is sampled by the error microphone through the linear superposition between $d(n)$ and $\widehat{y}(n)$ and expressed as
$e(n) = d(n) - y(n)\ast \bm s(n)$, 	where  $\bm s(n)$ represents the impulse response of the secondary path $S(z)$. In accordance with the compensation principle based on the effect of the secondary path $S(z)$ and slow adaptation, the residual noise $e(n)$ can be rewritten as
\begin{equation}
	e(n) = d(n) - \widehat{y}(n) = d(n) - \widehat{\mathcal{\bm A}}^\mathrm{T}(n)\bm \omega(n)
	\text{,}
	\label{058}
\end{equation}
where $\widehat{\mathcal{\bm A}}(n) = \mathcal{\bm A}(n)\ast\widehat{\bm s}(n)$ denotes the filtered expanded input vector, and $\widehat{\bm s}(n)$ is the impulse response of the estimated secondary path $\widehat{S}(z)$. Under the ergodicity assumption, the ensemble MSE can be approximated by the time average of the instantaneous squared errors. Using the instantaneous squared error as a stochastic estimate of the MSE and following (\ref{008}), the update of the FgLMS algorithm is given by
\begin{equation}
	\bm \omega(n+1) = \bm{\omega}(n) + \mu e(n)\widehat{\mathcal{\bm A}}(n)\text{.}
	\label{059}
\end{equation}

To evaluate the performance of the FgLMS algorithm, three NANC scenarios are considered. 
The VFxLMS \cite{tan2001adaptive}, FsLMS \cite{das2004active}, GFsLMS \cite{sicuranza2011generalized}, and AEFsLMS \cite{patel2016design} algorithms are used as the benchmarks. To evaluate the algorithms, the averaged noise reduction (ANR) is adopted as the performance index \cite{YIN2024110887}
\begin{equation}
	\text{ANR}(n)\text{,}\;\text{dB}= 20\log_{10}\left\{\frac{A_e(n)}{A_d(n)}\right\}\text{,}
	\label{068}
\end{equation}
where $A_e(n) = \lambda A_e(n-1) + (1-\lambda)\left|e(n)\right|$,
$A_d(n) = \lambda A_d(n-1) + (1-\lambda)\left|d(n)\right|$, and $\lambda=0.999$ denotes the forgetting factor.

In the first example, the logistic chaotic noise is exploited as the reference signal, which is expressed as \cite{YIN2024110887}
\begin{equation}
	x(n) = \kappa x(n-1)\left[1-x(n-1)\right]\text{,}
\end{equation}
where $\kappa=4$ and $x(0) = 0.9$. The primary path is  $P(z) = z^{-3}-0.3z^{-4}+0.2z^{-5}$, and the primary noise is generated by the third-order polynomial model
\begin{equation}
	\breve{d}(n) = \alpha(n-2)+c_1 \alpha^2(n-2) + c_2 \alpha^3(n-1)\text{,}
\end{equation}
where $c_1$ and $c_2$ are utilized to determine the nonlinearity of the primary path with $c_1=0.8$ and $c_2=-0.4$. $\alpha(n)$ represents the convolution of the reference noise $x(n)$ with the primary path with $\alpha(n) = x(n)\ast p(n)$ and $p(n)$ denotes the impulse response of the primary path $P(z)$. 

To evaluate the robustness of the FgLMS algorithm, the simulation incorporates an abrupt secondary path transition at iteration 100,000, switching from the minimum-phase secondary path $S(z) = z^{-2} + 0.5z^{-3}$ to the non-minimum-phase secondary path $S(z) = z^{-2} + 1.5z^{-3} - z^{-4}$ \cite{YIN2024110887}. Under the matched secondary path assumption, the estimated secondary path is switched synchronously to the corresponding non-minimum-phase secondary path for each algorithm. The observation noise has a variance of $0.0001$ and SNR $= 35.74$ dB.For all algorithms, the structures are adopted within the comparable dimensions and the parameters are selected to achieve the same convergence for a fair comparison during the initial minimum-phase stage with the VFxLMS algorithm ($N=25$, $L=74$, with $\mu$ adjusted from $3\times10^{-2}$ to $2.5\times10^{-2}$), the FsLMS algorithm ($N=15$, $L=75$, $\mathcal{B}=2$, with $\mu$ adjusted from $3\times10^{-2}$ to $5\times10^{-4}$), the GFsLMS algorithm ($N=7$, $L=59$, $\mathcal{B}=2$, $P=1$, with $\mu$ adjusted from $2\times10^{-2}$ to $2\times10^{-3}$), the AEFsLMS algorithm ($N=20$, $L=101$, $\mathcal{B}=2$, with $\mu$ adjusted from $8\times10^{-4}$ to $3\times10^{-4}$, and $\mu_a=2\times10^{-2}$), the FgLMS algorithm ($N=10$, $L=51$, $\mathcal{B}=2$, with $\mu$ adjusted from $8\times10^{-4}$ to $10^{-2}$, and $\gamma=0.2$).

\begin{figure}[htbp]
	\centering
	\begin{minipage}{0.49\linewidth}
		\centering
		\subfigure[]{\includegraphics[width=1.05\linewidth]{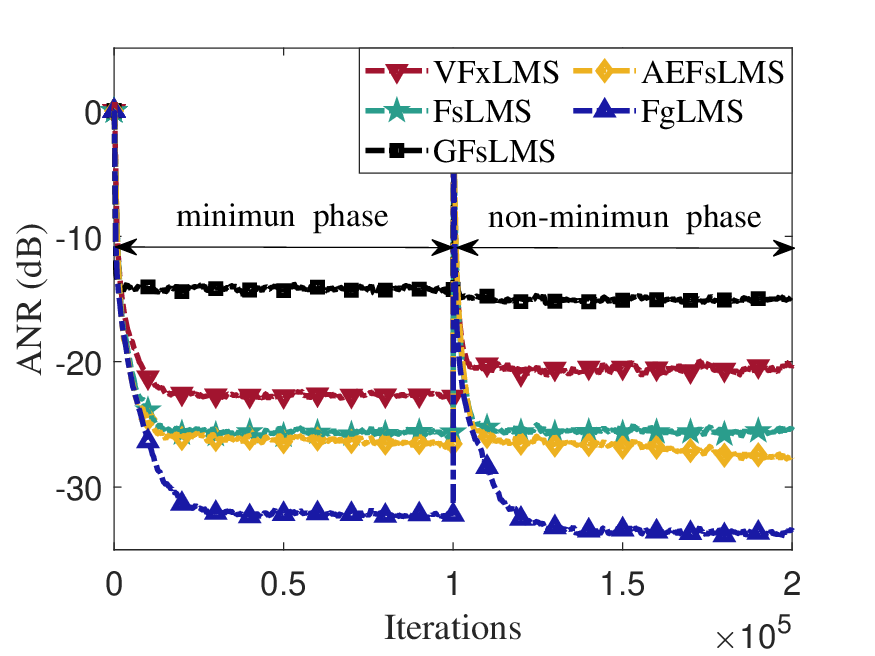}}
	\end{minipage}
	\begin{minipage}{0.49\linewidth}
		\centering
		\subfigure[]{\includegraphics[width=1.05\linewidth]{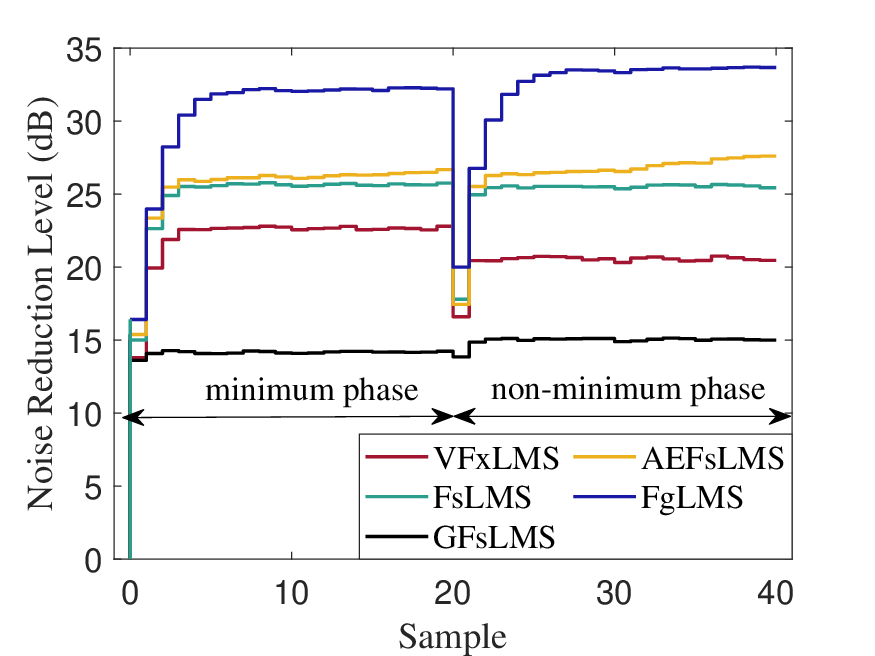}}
	\end{minipage}
	\caption{Comparison of the VFxLMS, FsLMS, GFsLMS, AEFsLMS, and FgLMS algorithms when the secondary path changes. (a) ANR. (b) Noise reduction levels.}
	\label{NANC_EX1}
\end{figure}

Fig. \ref{NANC_EX1}(a) shows the ANRs of the VFxLMS \cite{tan2001adaptive}, FsLMS \cite{das2004active}, GFsLMS \cite{sicuranza2011generalized}, AEFsLMS \cite{patel2016design}, and FgLMS algorithms, where the FgLMS algorithm achieves superior ANR performance, particularly under the non-minimum-phase condition. This improvement is attributed to its effective utilization of high-order input information for nonlinear system modeling. 

To further exhibit the effectiveness of the FgLMS algorithm, Fig. \ref{NANC_EX1}(b) presents the noise reduction levels averaged over intervals of 10,000 iterations under logistic chaotic noise \cite{10339836}. According to Fig. \ref{NANC_EX1}(b), the FgLMS algorithm maintains the highest level of noise reduction across both stages. Specifically, compared to the AEFsLMS algorithm, the FgLMS algorithm provides an additional noise reduction of approximately 10 dB during the minimum-phase stage observed at the 8th–9th interval and 7 dB during the non-minimum-phase stage observed at the 13th–14th interval.

In the second example, the nonlinear primary and secondary paths are used to validate the FgLMS algorithm. The input-output relationship of the nonlinear system is given by \cite{patel2016design}
\begin{equation}
	\breve{d}(n) = \text{tanh}\left\{3/\left\{1+\exp\left[-2x_f^2(n)\right]\right\}\right\}\text{,}
	\label{072}
\end{equation}
where $x_f(n) = 2x(n)/\left[1 + x^2(n)\right]$ with $x(n)\sim\mathcal{N}(0,0.1)$. 
The anti-noise $\widehat{y}(n)$ is obtained by the output $y(n)$ of the adaptive controller propagating through the nonlinear secondary path, where the nonlinear mapping is characterized as \cite{patel2016design}
\begin{equation}
	\begin{aligned}
		\widehat{y}(n) = &\;y(n) + 0.35y(n-1) + 0.9y(n-2)\\
		& - 0.15y(n)y(n-1) + 0.04y(n)y(n-2)\text{.}
		\label{073}
	\end{aligned}
\end{equation}

\begin{figure}[htbp]
	\centering
	\begin{minipage}{0.49\linewidth}
		\centering
		\subfigure[]{\includegraphics[width=1.05\linewidth]{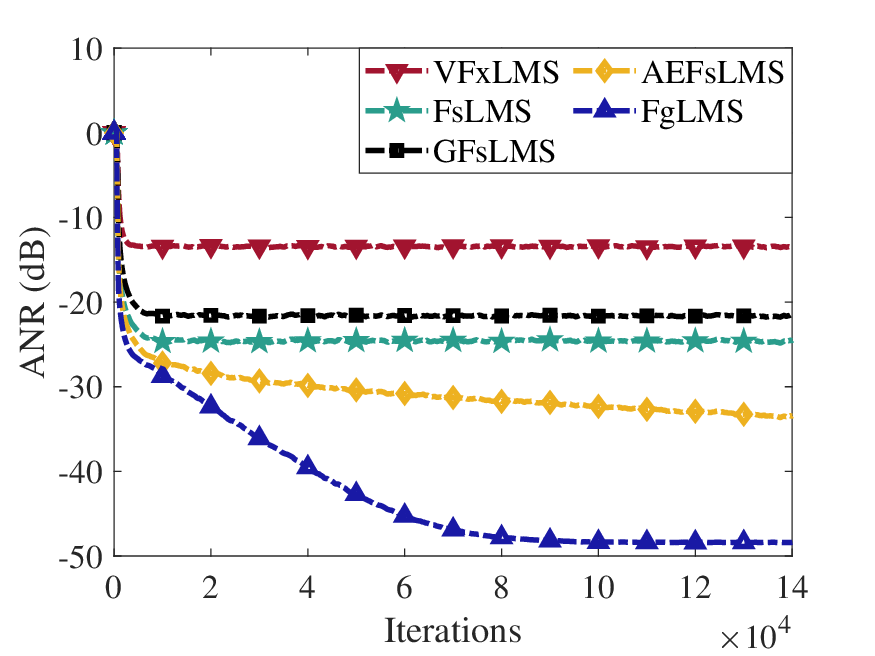}}
	\end{minipage}
	\begin{minipage}{0.49\linewidth}
		\centering
		\subfigure[]{\includegraphics[width=1.05\linewidth]{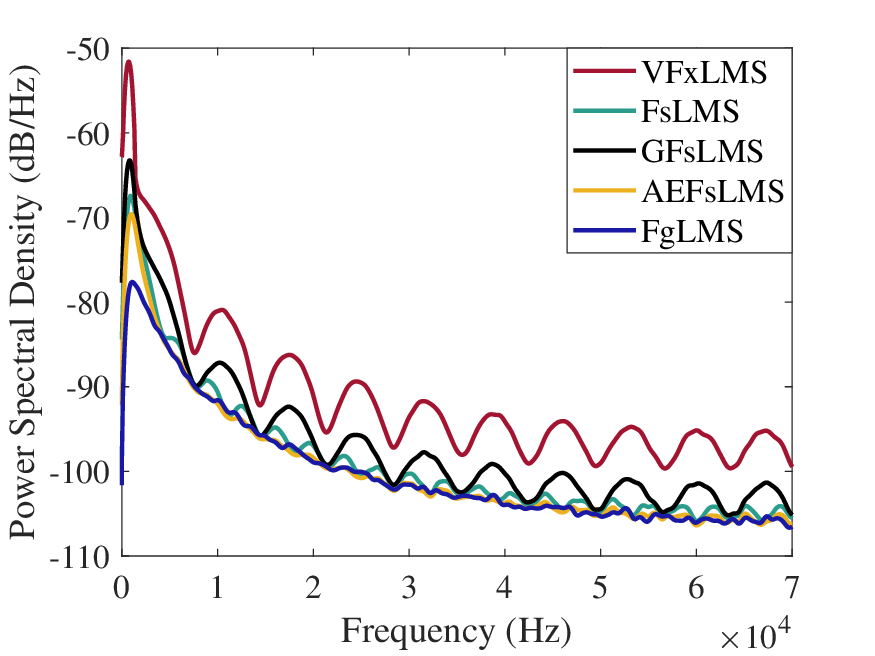}}
	\end{minipage}
	\caption{Comparison of the VFxLMS, FsLMS, GFsLMS, AEFsLMS, and FgLMS algorithms for the NANC scenario. (a) ANR. (b) PSD.}
	\label{NANC_EX2}
\end{figure}

To reach comparable convergence across all algorithms, the structures and step sizes are chosen as follows: the VFxLMS algorithm ($N=20$, $L=230$, and $\mu=2\times10^{-2}$), the FsLMS algorithm ($N=40$, $L=200$, $\mathcal{B}=2$, and $\mu=6\times10^{-4}$), the GFsLMS algorithm ($N=20$, $L=248$, $\mathcal{B}=2$, $P=2$, and $\mu=8\times10^{-4}$), the AEFsLMS algorithm ($N=40$, $L=201$, $\mathcal{B}=2$, $\mu=4\times10^{-4}$, and $\mu_a = 0.1$), and the FgLMS algorithm ($N=20$, $L=101$, $\mathcal{B}=2$, $\mu=5\times10^{-3}$, and $\gamma=1$). Fig. \ref{NANC_EX2}(a) shows the ANR of the FgLMS algorithm with the benchmark algorithms under observation noise with a variance of $0.0001$ and SNR $=30$ dB. One can see that the FgLMS algorithm can achieve an average ANR of $-48.39$ dB by averaging the final 10000 iterations. Remarkably, the FgLMS algorithm outperforms the VFxLMS and AEFsLMS algorithms by approximately 35 dB and 20 dB, respectively.

To evaluate the frequency-domain performance, Fig. \ref{NANC_EX2}(b) plots the power spectral density (PSD) of the algorithms,  computed using a 128-point Hann window. It can be seen that the FgLMS algorithm effectively suppresses low-frequency noise components within the 0–4200 Hz band. This enhanced noise reduction performance is attributed to the more accurate capture of the local information of reference noise.

In the third example, a 150000-sample traffic noise with a frequency range of 100 to 3500 Hz sampled at 8 kHz is utilized to evaluate the practical applicability of the FgLMS algorithm. The primary path is modeled by the polynomial nonlinear model, yielding the primary noise \cite{sicuranza2011generalized}
\begin{equation}
		\begin{aligned}
			\breve{d}(n) =&\; x(n) + 0.8x(n-1)+ 0.3 x(n-2) + 0.4x(n-3)\\
			& - 0.8x(n)x(n-1) + 0.9 x(n)x(n-2)\\
			&+ 0.7x(n)x(n-3)-3.9x^2(n-1)x(n-2)\\
			& - 2.6x^2(n-1)x(n-3) + 2.1 x^2(n-2)x(n-3)\text{.}
			\label{074}
		\end{aligned}
\end{equation}
The nonlinear secondary path is described by the SOV model and expressed as \cite{patel2016design}
\begin{equation}
	\begin{aligned}
		\widehat{y}(n) = &\;y(n) + 0.35 y(n-1) + 0.09 y(n-2)\\
		&-0.5y(n)y(n-1) + 0.4y(n)y(n-2)\text{.}
		\label{075}
	\end{aligned}
\end{equation}

\begin{figure}[!htbp]
	\centering
	\begin{minipage}{0.49\linewidth}
		\centering                             
		\subfigure[]{\includegraphics[width=1.05\linewidth]{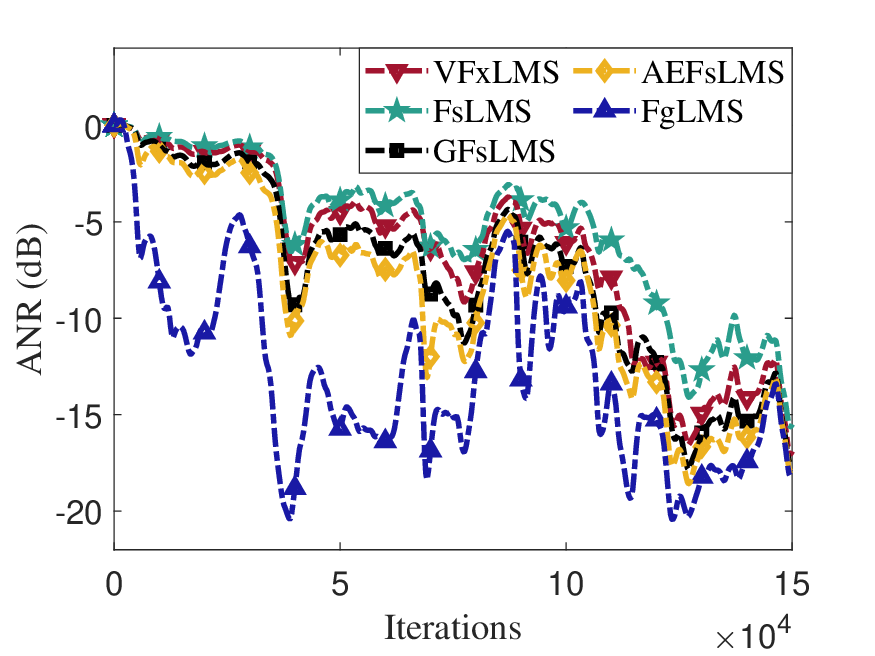}}
	\end{minipage}
	\begin{minipage}{0.49\linewidth}
		\centering
		\subfigure[]{\includegraphics[width=1.05\linewidth]{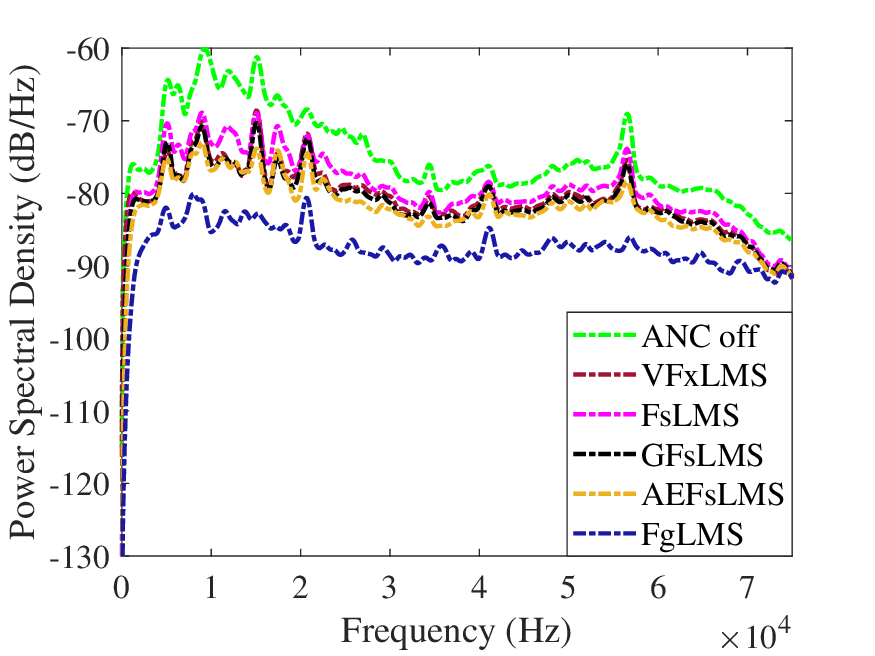}}
	\end{minipage}
	\caption{Comparison of the VFxLMS, FsLMS, GFsLMS, AEFsLMS, and FgLMS algorithms on the traffic noise. (a) ANR. (b) PSD.}
	\label{NANC_EX3}
\end{figure}

The functional expansion structures of the five algorithms are configured with the same settings as in the second example. To achieve the comparable convergence, the parameter settings are selected as VFxLMS ($\mu=5\times10^{-3}$), FsLMS ($\mu=8\times10^{-5}$), GFsLMS ($\mu=1.5\times10^{-4}$), AEFsLMS ($\mu=2\times10^{-4}$ and $\mu_a = 10^{-2}$), FgLMS ($\mu=2\times10^{-3}$ and $\gamma=2$). Fig. \ref{NANC_EX3}(a) shows the ANR of the FgLMS algorithm with the benchmarks. The rapid variations and high-frequency components of traffic noise induce tracking fluctuations across all algorithms. From Fig. \ref{NANC_EX3}(a), one can see that the FgLMS algorithm can achieve the smallest residual noise and the best noise reduction performance. To further illustrate the performance of the FgLMS algorithm, the PSDs of the five algorithms are shown in Fig. \ref{NANC_EX3}(b). Note that the primary traffic noise exhibits a broadband spectrum. Compared to the benchmarks suffering from degraded noise reduction performance, the FgLMS algorithm can eliminate the traffic noise with a broad frequency spectrum and accurately model the NANC system, yielding the improved noise reduction performance.

\section{Conclusion} 
\label{conclusion}
This work has proposed the GTFLN filter to enhance the localized nonlinear representation capability of TFLN-based filters while retaining a fixed-dimensional LIP structure. The theoretical analysis has established the advantages of the GTFLN filter over the AETFLN filter in terms of smoothness, RKHS, approximation error, and operator theory. The finite-dimensional approximation performance of the GTFLN filter has been analyzed jointly with the expansion order, tap length, and scaling parameter. Moreover, the steady-state EMSE analysis of the GTFLN filter has been presented, and the theoretical results have closely agreed with the simulated steady-state EMSEs. To reduce the sensitivity of the GTFLN filter, a locally optimized parameter update has been derived under the convergence condition, yielding the OGTFLN filter. The improved performance of the GTFLN and OGTFLN filters has been confirmed by the lower steady-state error compared with the SOV, TFLN, GeTFLN, and AETFLN benchmarks. Nevertheless, the OGTFLN filter provided a modest improvement over the GTFLN filter at the cost of increased computational complexity. Because the optimized scaling parameter update performs a local refinement around the nominal value, the GTFLN filter provides the tradeoff between the performance and complexity for resource-constrained applications. By contrast, the OGTFLN filter is suitable when the reduction in the steady-state error is required and the additional computational cost is acceptable.

The GTFLN filter has been equipped with the FgLMS algorithm for NANC and applied to NAEC. In both simulated and real-world noise environments, the FgLMS algorithm has demonstrated enhanced stability and better noise reduction performance compared to the VFxLMS, FsLMS, GFsLMS, and AEFsLMS algorithms. For NAEC under both single-talk and double-talk conditions, the GTFLN filter has exploited the GF to model the nonlinear loudspeaker echo path more accurately, thereby achieving improved echo attenuation over the benchmark nonlinear filters. The NANC experiments have been conducted under the matched secondary path condition. Investigating the effects of the secondary path mismatch, developing the online secondary path modeling, and integrating the GTFLN filter with robust NANC strategies for the impulsive noise environment are important directions for future work.

\appendix
\section*{Improvement of GTFLN Over AETFLN }
\subsection{Comparison of the Smoothness}
\label{appen1}
In approximation theory, the smoothness of a function $\mathcal{Y}:\mathbb{R}\to\mathbb{R}$ is characterized by its differentiability class. Here, $\mathbb{R}$ denotes the set of real numbers, $\mathit{C}^k(\mathbb{R})$ denotes the class of functions whose derivatives up to order $k$ are continuous, and $\mathit{C}^\infty(\mathbb{R})$ denotes the class of infinitely differentiable functions. The differentiability class affects the convergence behavior and remainder bounds of approximations, thereby influencing the estimation accuracy in nonlinear system modeling \cite{calderon1961lebesgue,JOHANSEN20001617}.

The adaptive exponential term in the AETFLN filter is given by $\varsigma\left[x(n)\right]=\exp\left[-a(n)\left|x(n)\right|\right]$ whose first derivative associated with $x(n)$ can be calculated with
\begin{equation}
	\varsigma{'}(n)=-a(n)\text{sign}\left[x(n)\right]\exp\left[-a(n)\left|x(n)\right|\right]\text{.}\label{A1}
\end{equation}
It is clear that when $x(n)=0$ and in a neighborhood of $0$, the right-hand derivative and left-hand derivative of $\varsigma\left[x(n)\right]=\exp\left[-a(n)\left|x(n)\right|\right]$ are calculated with $\lim\limits_{x(n)\to0^+}\varsigma{'}\left[x(n)\right] = \lim\limits_{x(n)\to0^+}\frac{\exp\left[-a(n)x(n)\right] -1}{x(n)}=-a(n)$ and $\lim\limits_{x(n)\to0^-}\varsigma{'}\left[x(n)\right] = \lim\limits_{x(n)\to0^-}\frac{\exp\left[a(n)x(n)\right] -1}{x(n)}=a(n)$, respectively. Hence, the first derivative $\varsigma{'}(n)$ is discontinuous, resulting in $\varsigma\left[x(n)\right]\in\mathit{C}^0\left(\mathbb{R}\right)$ but $\varsigma\left[x(n)\right]\notin\mathit{C}^1\left(\mathbb{R}\right)$.
Utilizing the Taylor series expansion in (\ref{016}), at $x(n)=0$ and in a neighborhood of $0$, $\varsigma\left[x(n)\right]$ can be expanded as
\begin{equation}
	\begin{aligned}
		\varsigma\left[x(n)\right]
		=&\;1 - a(n)\left|x(n)\right| + \frac{a^2(n)}{2}x^2(n)\\
		& -\frac{a^3(n)}{6}\left|x(n)\right|^3 + \mathrm{R}\left[x^4(n)\right]\text{.}
	\end{aligned}
	\label{A2}
\end{equation}
Due to the presence of $\left|x(n)\right|^k$ for odd $k$, $\varsigma\left[x(n)\right]$ is not real analytic and non-differentiable at the origin when $a(n)\neq0$. 

Utilizing the Taylor series expansion in (\ref{077}) yields
\begin{equation}
	\mathcal{G}\left[x(n)\right]
	= 1 - \gamma x^2(n) + \frac{\gamma^2}{2}x^4(n)-\frac{\gamma^3}{6}x^6(n) + \mathrm{R}\left[x^8(n)\right]\text{.}
	\label{A3}
\end{equation}
It follows that the GF $\mathcal{G}\left[x(n)\right]\in\mathit{C}^0(\mathbb{R})$. For the first derivative of $\mathcal{G}\left[x(n)\right]$, we have $\mathcal{G}'\left[x(n)\right]=\frac{\mathrm{d}}{\mathrm{d}x(n)}\exp\left[-\gamma x^2(n)\right]=-2\gamma x(n)\exp\left[-\gamma x^2(n)\right]$, where the whole term is continuous, resulting in $\mathcal{G}\left[x(n)\right]\in\mathit{C}^1(\mathbb{R})$. 

For the higher-order derivatives $\mathcal{G}^{(k)}\left[x(n)\right]$, we have $\mathcal{G}^{(k)}\left[x(n)\right]= \Theta_k\left[x(n)\right]\exp\left[-\gamma x^2(n)\right]$, where $\Theta_k\left[x(n)\right]$ is a polynomial of degree $k$. For
	$k=0$, we have $\Theta_0\left[x(n)\right]=1$. Suppose that this expression holds for some $k\geq 0$. Then, for the $(k+1)$th derivative, we have
	\begin{equation}
		\begin{aligned}
			&\mathcal{G}^{(k+1)}\left[x(n)\right]\\ 
			&= \frac{\mathrm{d}\left\{\Theta_k\left[x(n)\right]\exp\left[-\gamma x^2(n)\right]\right\}}{\mathrm{d}x(n)}\\
			&=\left\{\Theta_k'\left[x(n)\right]-2\gamma x(n)\Theta_k\left[x(n)\right]\right\}\exp\left[-\gamma x^2(n)\right]\text{.}
			\label{A4}
		\end{aligned}
	\end{equation}
	Let $\Theta_{k+1}\left[x(n)\right]=\Theta_k'\left[x(n)\right]-2\gamma x(n)\Theta_k\left[x(n)\right]$. Since $\Theta_k'\left[x(n)\right]$ has degree $k-1$ for
	$k\geq1$ and $\Theta_0'\left[x(n)\right]=0$ for $k=0$, whereas $x(n)\Theta_k\left[x(n)\right]$ has degree $k+1$, $\Theta_{k+1}\left[x(n)\right]$ is a polynomial of degree $k+1$. By mathematical induction,
	$\Theta_k\left[x(n)\right]$ is a polynomial of degree $k$ for every $k\geq0$.
	Hence, all derivatives of $\mathcal{G}\left[x(n)\right]$
	exist and are continuous, implying that $\mathcal{G}\left[x(n)\right]\in\mathit{C}^\infty(\mathbb{R})$.

Based on (\ref{A2}) and (\ref{A4}), $\varsigma\left[x(n)\right]$ belongs to $\mathit{C}^0(\mathbb{R})$ but not to $\mathit{C}^1(\mathbb{R})$. In contrast, since $\mathcal{G}\left[x(n)\right]$ is entire, its restriction to $\mathbb{R}$ is real analytic and belongs to $\mathit{C}^\infty(\mathbb{R})$, thereby facilitating high-order approximation and enhancing the modeling ability in characterizing complex nonlinear systems.

$\hfill\blacksquare$
\subsection{Comparison of the RKHS Property}
\label{appen2}
Based on the RKHS property, the GTFLN filter incorporating the GF demonstrates superior modeling capability compared to the AETFLN filter. The RKHS as a functional Hilbert space, denoted herein as $\mathcal{H}_K$, is characterized by the continuity and linearity of its point evaluation functionals, guaranteeing that for any function $f\in\mathcal{H}_K$ and any point $x$ in the input space $\mathcal{X}\subset\mathbb{R}$, $f(x)$ can be expressed as an inner product $f(x)=\left\langle f\text{,}\;\mathcal{K}(\cdot\text{,}\,x)\right\rangle_{\mathcal{H}_K}$, in which $\mathcal{K}:\mathcal{X}\times\mathcal{X}\to\mathbb{R}$ denotes the symmetric positive definite kernel \cite{berlinet2011reproducing,paulsen2016introduction}. Based on the \textit{Moore-Aronszajn} theorem \cite{aronszajn1950theory}, there is a bijective relationship between the positive definite kernel $\mathcal{K}$ and the RKHS. For $y=0$, the Gaussian kernel $\mathcal{K}_{\mathcal{G}}(x,y)=\exp[-|x-y|^2/(2\sigma^2)]$ and the Laplacian kernel $\mathcal{K}_{\mathcal{L}}(x,y)=\exp(-|x-y|/\theta)$ reduce to the GF and AEF for $\gamma=1/(2\sigma^2)$ and $a(n)=1/\theta$, respectively \cite{liu2011kernel}. These correspondences concern only their centered univariate profiles and the GF and AEF are not used as reproducing kernels. Because the two kernels induce different RKHSs, the Mat\'{e}rn RKHS family is adopted as a common comparison framework and defined as \cite{pmlr-v108-janz20a}
\begin{equation}
	\mathcal{K}_\mathcal{M}(x\text{,}\,y) = \frac{2^{1-\upsilon}}{\Gamma(\upsilon)}\left(\frac{\sqrt{2\upsilon}\left|x-y\right|}{l}\right)^\upsilon\mathcal{K}_\upsilon\left(\frac{\sqrt{2\upsilon}\left|x-y\right|}{l}\right)\text{,}
	\label{B1}
\end{equation}
where $\mathcal{K}_\upsilon(\cdot)$ represents the modified Bessel function of the second kind, $l$ is the length parameter, $\Gamma(\cdot)$ denotes the Gamma function, and $\upsilon>0$ is the smoothness parameter. When $\upsilon=1/2$ and $\upsilon\to\infty$, the RKHS associated with the Mat\'{e}rn kernel reduces to the same functional form, including the AEF and GF, respectively. In addition, the RKHS $\mathcal{H}_{K_\upsilon}$ associated with the Mat\'{e}rn kernel is equivalent to the Sobolev space $\mathcal{W}^{k\text{,}\;2}$ with $k=\upsilon + 1/2$ in the one-dimensional condition. More precisely, it refers to the equivalence of the norms when the Sobolev space is extended to $\mathbb{R}$.  Based on the Fourier transform \cite{sneddon1995fourier}, when the input signal dimension is $d=1$, the Fourier transform of the Mat\'{e}rn RKHS is given as
\begin{equation}
	\begin{aligned}
		\mathcal{M}_\upsilon(\omega)=\int_{-\infty}^{\infty}	\mathcal{K}_\mathcal{M}(\tau)e^{-i \omega\tau}\mathrm{d}\tau=\frac{2\sqrt{\pi}l(2\upsilon)^\upsilon\Gamma(\upsilon+1/2)}{\Gamma(\upsilon)\left(2\upsilon +l^2 \omega^2\right)^{\upsilon+1/2}}\text{.}
		\label{B2}
	\end{aligned}
\end{equation}
A function $f$ belongs to $\mathcal{H}_{K_\upsilon}$ if and only if its Fourier transform $\widehat{f}$ satisfies \cite{paulsen2016introduction}
\begin{equation}
	\left\|f\right\|^2_{\mathcal{H}_{K_\upsilon}} = \frac{1}{2\pi}\int_{\mathbb{R}}\frac{|\widehat{f}(\omega)|^2}{\mathcal{M}_\upsilon(\omega)}\mathrm{d}\omega <\infty\text{.}
	\label{B3}
\end{equation}

Note that a smoother RKHS contains a smaller class of the smooth functions, which can enhance the nonlinear modeling capability \cite{pmlr-v108-janz20a}. For comparison between $\mathcal{G}\left[x(n)\right]$ and $\varsigma\left[x(n)\right]$, the norms in the Mat\'{e}rn RKHS for both of them are calculated. For the AEF $\varsigma\left[x(n)\right]$, its Fourier transform is computed by
\begin{equation}
	\mathcal{F}_a(\omega)=\int_{-\infty}^{\infty}e^{-a(n)\left|x(n)\right|}e^{-i\omega x(n)}\mathrm{d}x(n) = \frac{2a(n)}{a^2(n) + \omega^2}\text{.}
	\label{B4}
\end{equation}
To determine whether $\varsigma\left[x(n)\right]$ belongs to $\mathcal{H}_{K_\upsilon}$, we analyze its Fourier transform $\widehat{f}_a(\omega)$, which is chosen as 
\begin{equation}
	\widehat{f}_a(\omega)= \frac{2a(n)}{a^2(n) + \omega^2}\text{.}
	\label{B5}
\end{equation}
Similarly, for $\mathcal{G}\left[x(n)\right]$, its Fourier transform is computed by
\begin{equation}
	\mathcal{F}_t(\omega)=\int_{-\infty}^{\infty}e^{-\gamma x^2(n)}e^{-i\omega x(n)}\mathrm{d}x(n)=\sqrt{\varepsilon}e^{-\frac{\omega^2}{4\gamma}}\text{,}
	\label{B6}
\end{equation}
where $\varepsilon=\pi/\gamma$.
Hence, to determine whether $\mathcal{G}\left[x(n)\right]$ belongs to $\mathcal{H}_{K_\upsilon}$, its Fourier transform $\widehat{f}_t(\omega)$ is taken as 
\begin{equation}
	\widehat{f}_t(\omega)= \sqrt{\varepsilon}e^{-\frac{\omega^2}{4\gamma}}\text{.}
	\label{B7}
\end{equation}

Considering a large $\upsilon$, the smooth RKHS approaches the Gaussian scenario. Taking $\upsilon=13/2$ as an example, although sample paths of a Mat\'{e}rn Gaussian process with $\upsilon=13/2$ are six times mean-square differentiable, the functions in the RKHS $\mathcal{H}_{K_\upsilon}$ are smoother, which is equivalent to the Sobolev Space $\mathcal{W}^{7\text{,}\;2}$. In this case, $\mathcal{M}_\frac{13}{2}(\omega)\sim(1+\omega^2)^{-7}$ for high-frequency attenuation.
Utilizing (\ref{B3}) and (\ref{B5}), the Mat\'{e}rn RKHS norm for the AEF $\varsigma\left[x(n)\right]$ can be expressed as
\begin{equation}
	\left\|f_a(\omega)\right\|^2_{\mathcal{H}_{K_\upsilon}}=\frac{1}{2\pi}\int_{\mathbb{R}}\frac{1}{\mathcal{M}_\upsilon(\omega)}\left|\frac{2a(n)}{a^2(n) + \omega^2}\right|^2\mathrm{d}\omega\text{,}
	\label{B8}
\end{equation}
Based on (\ref{B8}), as $\omega\to\infty$, the integrand behaves asymptotically as
\begin{equation}
	\frac{4a^2(n)}{\left[a^2(n) + \omega^2\right]^2}\cdot(1 + \omega^2)^7\sim\frac{1}{\omega^4}\cdot\omega^{14} = \omega^{10}\to\infty\text{.}
	\label{B9}
\end{equation}
The asymptotic behavior in (\ref{B9}) shows that the integral diverges for $\upsilon=13/2$. As $\omega\to\infty$, the Fourier transform of $\varsigma\left[x(n)\right]$ decays algebraically as $1/\omega^2$. Because $\vert\widehat{f}_a(\omega)\vert^2\sim1/\omega^4$ and $\mathcal{M}_\upsilon^{-1}(\omega)\sim\omega^{2\upsilon+1}$, the integrand in (\ref{B3}) behaves as $\omega^{2\upsilon-3}$, which is not integrable for $\upsilon\geq1$. Therefore, $\varsigma\left[x(n)\right]\notin\mathcal{H}_{K_\upsilon}$ for $\upsilon\geq1$. Similarly, using (\ref{B3}) and (\ref{B7}), the Mat\'{e}rn RKHS norm of the GF $\mathcal{G}\left[x(n)\right]$ is given by
\begin{equation}
	\left\|f_t(\omega)\right\|^2_{\mathcal{H}_{K_\upsilon}}=\frac{1}{2\pi}\int_{\mathbb{R}}\frac{1}{\mathcal{M}_\upsilon(\omega)}\left|\sqrt{\varepsilon}e^{-\frac{\omega^2}{4\gamma}}\right|^2\mathrm{d}\omega\text{,}
	\label{B10}
\end{equation}
Based on (\ref{B10}), as $\omega\to\infty$, the integrand behaves asymptotically as
\begin{equation}
	\varepsilon e^{-\frac{\omega^2}{2\gamma}}\cdot(1 + \omega^2)^7\sim e^{-\frac{\omega^2}{2\gamma}}\cdot\omega^{14} = \omega^{14}/e^{\frac{\omega^2}{2\gamma}}\to 0\text{.}
	\label{B11}
\end{equation}
This result in (\ref{B11}) indicates that $e^{-\frac{\omega^2}{2\gamma}}$ decays significantly faster than $\omega^{14}$ grows. Hence, the integrand decays exponentially, ensuring convergence of the integral. One can see that for all $\upsilon>0$, $\mathcal{G}\left[x(n)\right]$ belongs to the $\mathcal{H}_{K_\upsilon}$.

Based on the Fourier domain norm formula and utilizing (\ref{B8}), we observe that the decay of the Fourier transform $\widehat{f}_a(\omega)$ for $\varsigma\left[x(n)\right]$ leads to the convergence of $\left\|f_a(\omega)\right\|^2_{\mathcal{H}_{K_\upsilon}}$, which requires the integrand $\left[a^2(n) + \omega^2\right]^{-2}\cdot\left(1 + \omega^2\right)^{\upsilon + 1/2}$ to be integrable. This condition is satisfied when $-4 + 2\upsilon +1 <-1$. Hence, we have $\upsilon<1$. In fact, the Mat\'{e}rn RKHS $\mathcal{H}_{K_\upsilon}$ is equivalent to the Sobolev space $\mathcal{W}^{k\text{,}\;2}(\mathbb{R})$ with $k = \upsilon + 1/2$. However, the AEF $\varsigma\left[x(n)\right]=\exp\left[-a(n)\left|x(n)\right|\right]$ belongs to $\mathcal{W}^{s\text{,}\;2}(\mathbb{R})$ for any $s<1.5$. Specifically, it is in $\mathcal{W}^{1\text{,}\;2}(\mathbb{R})$ but not $\mathcal{W}^{2\text{,}\;2}(\mathbb{R})$ and as such, $k<1.5$ is equivalent to $\upsilon+1/2<1.5$, which confirms the result $0<\upsilon<1$.
Exploiting (\ref{B10}), the Fourier transform $\widehat{f}_t(\omega)$ for $\mathcal{G}\left[x(n)\right]$ exhibits the exponential decay compared to that of the Mat\'{e}rn RKHS with polynomial decay, resulting in the fact that $\mathcal{G}\left[x(n)\right]$ belongs to RKHS for all $\upsilon>0$. As a result, the GTFLN filter can effectively model the unknown system with the enhanced modeling capability compared to the AETFLN filter.
$\hfill\blacksquare$
\subsection{Comparison of the Approximation Error}
\label{theory3}
Based on the Bernstein theorem \cite{schilling2012bernstein}, suppose that a function $f(x)$ defined on the interval $\left[-1\text{,}\; 1\right]$ admits an analytic continuation to the interior of the Bernstein ellipse $\varepsilon_\rho$ with the foci $\pm $1 and parameter $\rho>1$ and is defined by
\begin{equation}
	\varepsilon_\rho=\left\{z\in\mathbb{C}:\left|z-1\right|+\left|z + 1\right|< \rho + \rho^{-1}\right\}\text{.}
	\label{C01}
\end{equation}
In addition, the analytic continuation $f(z)$ of $f(x)$ meets $\left|f(z)\right|\leq M$. The best uniform approximation error $\mathcal{E}_k(f)$ by the polynomials ${P}_k$ with degree $k$ satisfies \cite{8332514,256500}
\begin{equation}
	\begin{aligned}
		\mathcal{E}_k(f) &= \inf\limits_{P_k\in\mathcal{P}_k}\left\|f - P_k\right\|\\
		&= \inf\limits_{P_k\in\mathcal{P}_k}\max\limits_{x\in[-1\text{,}\;1]}\left|f(x) - P_k(x)\right|\leq \frac{2M}{\rho^k\left(\rho - 1\right)}
		\label{C02}
	\end{aligned}
\end{equation}

According to Appendix \ref{appen1}, $\mathcal{G}\left[x(n)\right]$ is entire with the analytic continuation $\mathcal{G}(z)=\exp\left(-\gamma z^2\right)$. The constrained problem for obtaining the maximum modulus $M(\rho)$ of $\mathcal{G}(z)$ can be calculated as
\begin{equation}
	\begin{aligned}
		M(\rho) &= \max\limits_{z\in \varepsilon_\rho}\left|\mathcal{G}(z)\right|= \max\limits_{z\in \varepsilon_\rho}\exp\left[-\gamma(x^2 - y^2)\right] \\
		&\;\text{s.t.} \quad\frac{x^2}{a_\rho^2} + \frac{y^2}{b_\rho^2} =1\text{,}
		\label{C03}
	\end{aligned}
\end{equation}
where $a_\rho=(\rho + \rho^{-1})/2$ and $b_\rho=(\rho - \rho^{-1})/2$. The minimum of $x^2-y^2$ is $-b_\rho^2$, attained at $x=0$ and $y=\pm b_\rho$. Therefore, the maximum modulus $M(\rho)$ is
\begin{equation}
	M(\rho) = \exp\left(\gamma b_\rho^2\right)=\exp\left[\gamma\left(\frac{\rho - \rho^{-1}}{2}\right)^2\right]\text{.}
	\label{C04}
\end{equation}

Substituting (\ref{C04}) into (\ref{C02}), the best uniform approximation error $\mathcal{E}_k\left(\mathcal{G}\right)$ concerning the GF $\mathcal{G}\left[x(n)\right]$ is expressed as 
\begin{equation}
	\mathcal{E}_k\left(\mathcal{G}\right)\leq 2\exp\left[\frac{\gamma}{4}\left(\rho - \rho^{-1}\right)^2\right]\rho^{-k}(\rho - 1)^{-1}\text{.}
	\label{C05}
\end{equation}

Since $\mathcal{G}\left[x(n)\right]$ is entire, (\ref{C05}) holds for any $\rho>1$. The optimal upper bound can be achieved by optimizing the parameter $\rho$. The optimization problem is considered with the approximation $\left(\rho - \rho^{-1}\right)^2\approx\rho^2$
\begin{equation}
	\min\limits_{\rho}\;\Xi(\rho) = \min\limits_{\rho}\; 2\exp\left(\frac{\gamma\rho^2}{4}\right)\cdot\rho^{-k}\text{.}
	\label{C06} 
\end{equation}
Taking the logarithm on both sides of (\ref{C06}) and utilizing the gradient descent method with respect to $\rho$ result in
\begin{equation}
	\frac{\partial \ln\Xi(\rho)}{\partial \rho} = \frac{\gamma\rho}{2} - \frac{k}{\rho}\text{.}
	\label{C07} 
\end{equation}
Setting $\frac{\partial \ln\Xi(\rho)}{\partial \rho}=0$ and solving for $\rho$ gives
\begin{equation}
	\rho = \sqrt{2k/\gamma}\text{.}
	\label{C08} 
\end{equation}

Employing the Stirling approximation and substituting (\ref{C08}) into (\ref{C05}) leads to 
\begin{equation}
	\mathcal{E}_k\left(\mathcal{G}\right)<2\exp\left(\gamma/4\cdot2k/\gamma\right)\cdot\left(\sqrt{2k/\gamma}\right)^{-k} \approx 2/\sqrt{k!}\cdot(\gamma/2)^{\frac{k}{2}}\text{.}
	\label{C09}
\end{equation}
	
	Based on (\ref{C09}), we have $\mathcal{E}_k\left(\mathcal{G}\right)\sim \mathcal{O}\left(\frac{1}{\sqrt{k!}}\right)$, where $\mathcal{G}\left[x(n)\right]$ satisfies the super-exponential convergence. 
	
	Because $\varsigma\left[x(n)\right]\in\mathit{C}^0(\mathbb{R})$ but $\varsigma\left[x(n)\right]\notin\mathit{C}^1(\mathbb{R})$ for $a(n)\neq0$, its polynomial approximation error is governed by the cusp at the origin. From (\ref{A2}), the leading nonsmooth term is $-a(n)\left|x(n)\right|$, whereas the remaining higher-order terms are smoother and have faster-decaying approximation errors. Therefore, the leading asymptotic contribution to the best uniform approximation error is
	\begin{equation}
		\mathcal{E}_k\left(\varsigma\right)\approx \left|a(n)\right|\mathcal{E}_k\left[\left|x(n)\right|\right]
		\text{.}
		\label{C10}
	\end{equation}
	Utilizing the approximation for the Bernstein constant $\beta\approx\frac{1}{2\sqrt{\pi}}$, the celebrated limit theorem states that $\lim\limits_{k\to\infty}k\mathcal{E}_k\left[\left|x(n)\right|\right]=\frac{1}{2\sqrt{\pi}}$ \cite{schilling2012bernstein}. Hence, the best uniform approximation error $\mathcal{E}_k\left(\varsigma\right)$ can be calculated as 
	\begin{equation}
		\mathcal{E}_k\left(\varsigma\right) \approx \left|a(n)\right|/(2\sqrt{\pi}k)\text{.}
		\label{C11}
	\end{equation}
	
	Based on (\ref{C11}), we have $\mathcal{E}_k\left(\varsigma\right)\sim \mathcal{O}(\frac{1}{k})$, where $\varsigma\left[x(n)\right]$ satisfies the algebraic convergence. 
	
	According to (\ref{C09}) and (\ref{C11}), the superiority of $\mathcal{G}\left[x(n)\right]$ over $\varsigma\left[x(n)\right]$ in terms of approximation error originates from the disparity between their functional analyticity and the resulting asymptotic convergence rates. According to Bernstein's theorem, the entire analyticity of $\mathcal{G}(z)=\exp\left(-\gamma z^2\right)$, which is devoid of singularities in the complex plane, guarantees a super-exponential decay of the best uniform approximation error at a rate of $\mathcal{O}(\frac{1}{\sqrt{k!}})$. However, the non-differentiable $\left|x(n)\right|$ term restricts $\varsigma\left[x(n)\right]$ to the class $\mathit{C}^0(\mathbb{R})$, yielding a slower algebraic convergence of $\mathcal{O}(\frac{1}{k})$. This theoretical disparity implies that for nonlinear system modeling, the $\mathcal{G}\left[x(n)\right]$-based GTFLN filter can achieve comparable estimation accuracy with a lower model order $k$ compared to its $\varsigma\left[x(n)\right]$-based AETFLN counterpart. This superior approximation performance translates into enhanced tractability and reduced computational complexity. Moreover, by achieving high-precision fitting with fewer coefficients, the $\mathcal{G}\left[x(n)\right]$-based architecture effectively mitigates the numerical ill-conditioning and overfitting risks in high-order expansions.
	$\hfill\blacksquare$
	
	\subsection{Comparison of the Operator Theory}
	\label{theory4}
	The Gaussian function $\mathcal{G}(x)=\exp\left(-\gamma x^2\right)$ is proportional, after appropriate scaling, to the ground-state wavefunction of the quantum harmonic oscillator (QHO) \cite{6375857}. The higher-order QHO eigenfunctions are Hermite--Gaussian functions that form a complete orthonormal basis of $L^2(\mathbb{R})$. Moreover, translated and scaled Gaussian functions can be used as basis functions in Gaussian RBF networks, whose finite linear combinations possess the universal approximation property \cite{392253}. By contrast, the Laplacian function $\varsigma(x)=\exp\left(-a|x|\right)$ is, up to a normalization factor, the Green's function of a modified Helmholtz operator and is associated with a singular delta term at the origin.
	
	Consider the self-adjoint Schr\"odinger operator $\widehat{\mathcal{H}}$ on the Hilbert space $L^2(\mathbb{R})$ \cite{6189036}
	\begin{equation}
		\widehat{\mathcal{H}} = -\frac{\mathrm{d}^2}{\mathrm{d}x^2} + x^2\text{.}
		\label{d01}
	\end{equation}
	Note that (\ref{d01}) provides the spectral decomposition of the operator, yielding the discrete spectrum of eigenvalues and a complete set of orthonormal eigenfunctions, such as Hermite functions with $\psi_i(x) = \frac{1}{\sqrt{2^i i! \sqrt{\pi}}} H_i(x) e^{-x^2/2}$, where $H_i(x)$ stands for the Hermite polynomials and when $i=0$, $\psi_0(x)$ reduces to the Gaussian function. 
	
	Applying the self-adjoint Schr\"odinger operator $\widehat{\mathcal{H}}$ to the harmonic oscillator eigenfunctions $\psi_i(x)$ yields \cite{6375857}
	\begin{equation}
		\widehat{\mathcal{H}}\psi_i(x) = \left[-\frac{\mathrm{d}^2}{\mathrm{d}x^2} + x^2\right]\psi_i(x) = (2i + 1)\psi_i(x)\text{.}
		\label{d03}
	\end{equation}
	Note that (\ref{d03}) guarantees that the basis $\left[\psi_i(x)\right]_{i=0}^\infty$ constitutes a complete orthonormal basis. The ground state wavefunction $\psi_0(x)$ can be given by $\psi_0(x)=\pi^{-1/4}\exp\left(-x^2/2\right)$. The relationship between  $\mathcal{G}\left[x(n)\right]$ and $\psi_0(x)$ is given by
	\begin{equation}
		\mathcal{G}\left[x(n)\right] = \pi^{1/4}\psi_0\left[\sqrt{2\gamma}x(n)\right]\text{.}
		\label{d04}
	\end{equation}
	Eq. (\ref{d04}) ensures that the GF is proportional to the ground state wavefunction of the harmonic oscillator, which provides the linear approximation for the nonlinear system based on the orthogonal basis. 
	
	Compared to the Gaussian function, the Laplacian kernel $\varsigma(x)=\exp\left(-a \left|x\right|\right)$ is not the eigenfunction of a smooth potential. However, it arises as the Green's function of the linear differential operator associated with the modified Helmholtz equation and is given by
	\begin{equation}
		\begin{aligned}
			\left(a^2 - \frac{\mathrm{d}^2}{\mathrm{d}x^2}\right)
			\varsigma(x) =a^2\varsigma(x) - \varsigma^{\prime\prime}(x)= 2a\delta(x)\text{,}
			\label{d05}
		\end{aligned}
	\end{equation}
	where $\varsigma^{\prime\prime}(x)=a^2\varsigma(x)-2a\delta(x)$ and $\delta(\cdot)$ stands for the Dirac delta function. 
	
	Based on (\ref{d03}) and (\ref{d05}), $\mathcal{G}\left[x(n)\right]$ outperforms $\varsigma\left[x(n)\right]$ due to its intrinsic regularity as the smooth ground state of a harmonic Schr\"odinger operator, contrasted with the singular delta-potential origin of the Laplacian kernel. The $\mathcal{G}\left[x(n)\right]$-based basis constitutes a complete orthonormal subspace with high-order differentiability. In contrast, $\varsigma\left[x(n)\right]$ corresponds to the bound state of a singular attractive delta potential and exhibits first-derivative discontinuities that limit approximation efficiency and trigger numerical instabilities in nonlinear fitting. Consequently, the $\mathcal{G}\left[x(n)\right]$-based architecture of GTFLN filter provides a more robust and efficient framework for capturing complex nonlinear manifolds compared to the singular $\varsigma\left[x(n)\right]$ basis.
	$\hfill\blacksquare$

	\ifCLASSOPTIONcaptionsoff
	\newpage
	\fi
	
	\footnotesize
	\bibliographystyle{IEEEtran}
	\bibliography{IEEEabrv,reference}
\end{document}